\documentclass[12pt]{cernart}
\usepackage{epsfig}
\usepackage{amsmath}
\usepackage{amssymb}
\usepackage{times}
\usepackage{multirow}
\begin{document}
 
 
 
\setcounter{topnumber}{3}
\renewcommand{\topfraction}{0.999}
\renewcommand{\bottomfraction}{0.99}
\renewcommand{\textfraction}{0.0}
\setcounter{totalnumber}{6}
 
\def\thefootnote{\fnsymbol{footnote}}
\begin{titlepage}
%
\title{\large{Inclusive production of charged pions in p+C collisions at 158 GeV/c \\
 beam momentum}}
%
\begin{Authlist}
\vspace{2mm}

%
%
%
%
%
%
\noindent
C.~Alt$^{7}$,
B.~Baatar$^{6}$, D.~Barna$^{3}$, G.~Barr$^{10}$, J.~Bartke$^{4}$, 
L.~Betev$^{8}$, H.~Bia{\l}\-kowska$^{13}$, C.~Blume$^{7}$,
B.~Boimska$^{13}$, J.~Bracinik$^{2}$, R.~Bramm$^{7}$, P.~Bun\v{c}i\'{c}$^{8}$, 
V.~Cerny$^{2}$,  P.~Christakoglou$^{1}$, O.~Chvala$^{11}$,
P.~Dinkelaker$^{7}$, J.~Dolejsi$^{11}$, 
V.~Eckardt$^{9}$, 
H.G.~Fischer$^{8}$, D.~Flierl$^{7}$, Z.~Fodor$^{3}$, P.~Foka$^{5}$, V.~Friese$^{5}$,
M.~Ga\'zdzicki$^{7}$, G~.Georgopoulos$^{1}$, 
C.~H\"{o}hne$^{5}$, 
A.~Karev$^{8,9}$, S.~Kniege$^{7}$, T.~Kollegger$^{7}$, 
V.I.~Kolesnikov$^{6}$, E.~Kornas$^{4}$, M.~Kowalski$^{4}$,
I.~Kraus$^{5}$, M.~Kreps$^{2}$, 
L.~Litov$^{12}$,
M.~Makariev$^{12}$, A.I.~Malakhov$^{6}$, M.~Mateev$^{12}$, 
G.L.~Melkumov$^{6}$, M.~Mitrovski$^{7}$,
G.~P\'{a}lla$^{3}$, A.D.~Panagiotou$^{1}$, D.~Panayotov$^{12}$, 
C.~Pattison$^{10}$, A.~Petridis$^{1}$,
R.~Renfordt$^{7}$, A.~Rybicki$^{4}$, 
A.~Sandoval$^{5}$, N.~Schmitz$^{9}$, P.~Seyboth$^{9}$, 
F.~Sikl\'{e}r$^{3}$, R.~Stock$^{7}$,
H.~Str\"{o}bele$^{7}$, J.~Sziklai$^{3}$, P.~Szymanski$^{8,13}$,
V.~Trubnikov$^{13}$, 
D.~Varga$^{8}$, M.~Vassiliou$^{1}$, G.I.~Veres$^{3}$, 
G.~Vesztergombi$^{3}$, D.~Vrani\'{c}$^{5}$,
S.~Wenig$^{8}$\footnote{Corresponding author: Siegfried.Wenig@cern.ch},
A.~Wetzler$^{7}$,
J.~Zaranek$^{7}$
\vspace*{2mm}
 
\noindent
{\it (The NA49 Collaboration)}  \\
\vspace*{2mm}
\noindent
$^{1}$Department of Physics, University of Athens, Athens, Greece.\\
$^{2}$Comenius University, Bratislava, Slovakia\\
$^{3}$KFKI Research Institute for Particle and Nuclear Physics, Budapest, Hungary\\
$^{4}$The H. Niewodnicza\'nski Institute of Nuclear Physics, 
      Polish Academy of Sciences, Cracow, Poland \\
$^{5}$Gesellschaft f\"{u}r Schwerionenforschung (GSI), Darmstadt, Germany.\\
$^{6}$Joint Institute for Nuclear Research, Dubna, Russia\\
$^{7}$Fachbereich Physik der Universit\"{a}t, Frankfurt, Germany\\
$^{8}$CERN, Geneva, Switzerland\\
$^{9}$Max-Planck-Institut f\"{u}r Physik, Munich, Germany\\
$^{10}$University of Oxford, Oxford, UK \\
$^{11}$Institute of Particle and Nuclear Physics, Charles University, Prague, Czech Republic\\
$^{12}$Atomic Physics Department, Sofia University St. Kliment Ohridski, Sofia, Bulgaria\\
$^{13}$Institute for Nuclear Studies, Warsaw, Poland\\
\end{Authlist}

\vspace*{2mm}

\begin{abstract}
\vspace{-3mm}
The production of charged pions in minimum bias p+C interactions is studied
using a sample of 377~000 inelastic events obtained with the NA49 detector
at the CERN SPS at 158~GeV/c beam momentum. The data cover a phase space
area ranging from 0 to 1.8~GeV/c in transverse momentum and from -0.1
to 0.5 in Feynman x. Inclusive invariant cross sections are given on
a grid of 270 bins per charge thus offering for the first time a dense
coverage of the projectile hemisphere and of the cross-over region into
the target fragmentation zone.
\end{abstract}
 
\cleardoublepage
 
\end{titlepage}

%
%
\section{Introduction}
\vspace{3mm}
\label{sec:intro}
The study of p+A interactions represents an important part of the NA49
experimental programme which is aimed at a comprehensive study of soft
hadronic interactions covering both elementary and nuclear collisions.
In the framework of this programme a first paper on inclusive pion 
production in p+p collisions has been published recently \cite{bib:pp_paper}.
The present study extends these measurements to p+C interactions. The use 
of the isoscalar, light $^{12}$C nucleus has several motivations. Firstly,
it allows inspection of the evolution from elementary to nuclear
collisions for a small number of intranuclear projectile interactions
as compared to the high-statistics data sets on the heavy $^{208}$Pb nucleus
from the NA49 experiment \cite{bib:hgf}. Secondly, it satisfies, at 
least partially, the need of high precision p+A reference data from 
light nuclei for the control of systematic effects in neutrino physics. This
applies both to long baseline neutrino oscillation experiments using
light neutrino production targets (e.g. \cite{bib:minos} which uses a beam 
energy of 120 GeV) and to atmospheric neutrino
studies \cite{bib:engel}. In the latter case the $^{12}$C nucleus is 
sufficiently close to $^{14}$N and $^{16}$O to allow a sensible test of 
the different hadronic production models developed for the interpretation 
of these data \cite{bib:neut}. 
 
The NA49 p+C data sample consists of 377~000 inelastic events corresponding 
to only 8\% of the event number available in p+p collisions 
\cite{bib:pp_paper}. The reasons for this relatively limited statistics 
lie entirely in the heavy restrictions which the SPS fixed-target programme 
as a whole has suffered over the past years. These restrictions concern 
both the beam availability and the choice of topics in hadronic physics 
acceptable to the relevant committee \cite{bib:prop}. In view of the almost total 
absence of existing data in the SPS energy range \cite{bib:barton,bib:niki} 
the NA49 results nevertheless constitute a unique extension of the present 
knowledge in the sector of p+A interactions with light nuclei. 

The present study is closely related to the preceding
publication on pion production in p+p interactions \cite{bib:pp_paper}
which is recommended as a reference for the description of most experimental 
details. Only those items which are specific for p+C collisions will be 
touched upon here. The layout of the paper is arranged as follows. 
After a short comment concerning the situation of previous measured data 
in Sect.~2, Sect.~3 describes the parts of the NA49 experiment which differ
from the p+p data taking. Sect.~4 deals with the determination of the
inclusive cross sections and the applied corrections. The final data
are presented  and compared to other experiments in Sects.~5 and 6. The
$p_T$ integrated distributions are evaluated for minimum bias condition 
in Sect.~7 and Sect.~8 describes the dependence on the measured number of
grey protons. 

%
%
\section{The Experimental Situation}
\vspace{3mm}
\label{sec:exp_sit}
The asymmetric nature of p+A interactions necessitates in principle
a complete coverage of the target and projectile hemispheres in
order to experimentally constrain the underlying production mechanism.
The phase space coverage of exisiting data in p+C
is even more restricted than for p+p collisions \cite{bib:pp_paper}. 
In fact there are only two sets of data in the SPS energy range offering
double differential inclusive cross sections for identified pions,

\begin{equation}
  \frac{d^2\sigma}{dx_Fdp_T^2}   ,
\end{equation}
where the longitudinal scaling variable $x_F = 2p_L/\sqrt{s}$ is defined 
in the nucleon-nucleon cms.
 
\begin{figure}[t]
  \centering
  \epsfig{file=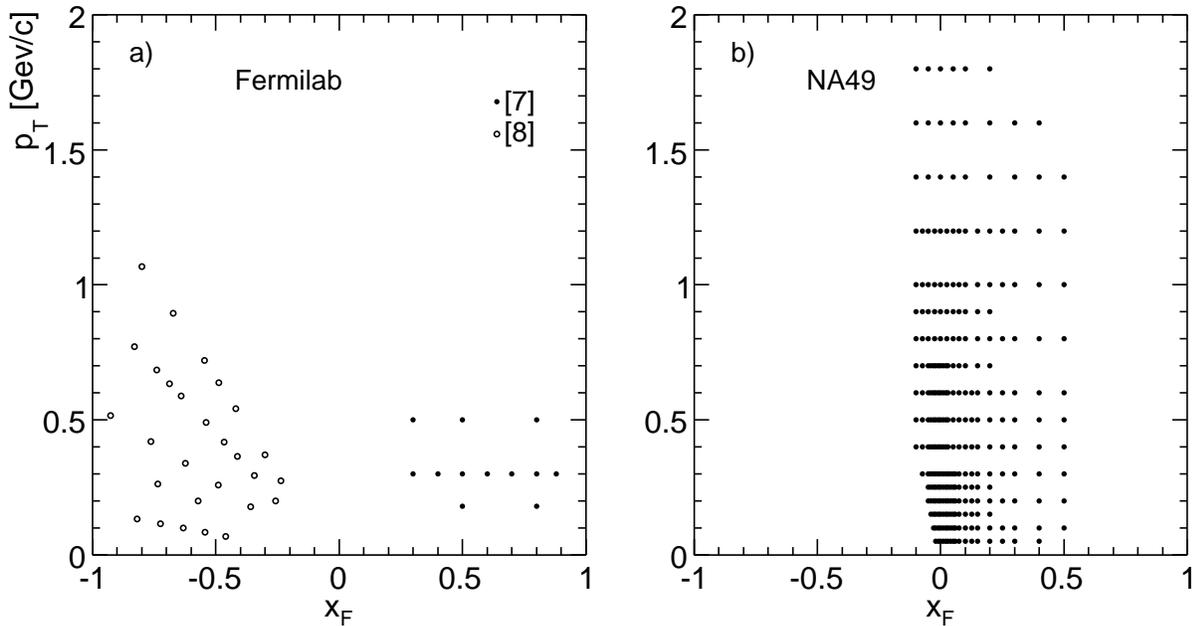,width=16cm}
  \caption{Phase space coverage of existing data in a) Fermilab and b) NA49}
  \label{fig:cov}
\end{figure}

The first data set at 400~GeV/c beam momentum covers the far backward
region at a number of fixed laboratory angles \cite{bib:niki}, yielding
important information about the region of intra-nuclear cascading.
The second one presents a small number of data points in the far forward
region \cite{bib:barton}. The situation is depicted in Fig.~\ref{fig:cov}a 
which shows that there are no data in the complete central region
-0.2~$<x_F<$~0.3.

The new NA49 results fill this gap to a large extent, as shown
in Fig.~\ref{fig:cov}b. They are limited to $p_T<$~1.8~GeV/c and to 
$x_F<$~0.5 essentially by the modest event statistics, see Sect.~\ref{sec:intro}. 
In the backward hemisphere, a $p_T$ dependent cut at $x_F>$~-0.1 is imposed 
both by the NA49 acceptance and by the particle identification via ionization 
energy loss. Nevertheless an inspection of the important cross-over region 
between projectile and target hemispheres is possible.

%
%
\section{Parts of the NA49 experiment specific to p+C interactions}
\vspace{3mm}

Most parts of the NA49 experimental setup and of the data processing 
procedure are identical for p+p and p+A data taking. This concerns 
the beam definition, the interaction trigger scheme, tracking, event 
reconstruction and selection, and particle identification \cite{bib:pp_paper}. 
The experimental items which are specific to the p+C data taking are 
described below.

%
%
\subsection{Target and grey proton detection}
\begin{figure}[h]
  \centering
  \epsfig{file=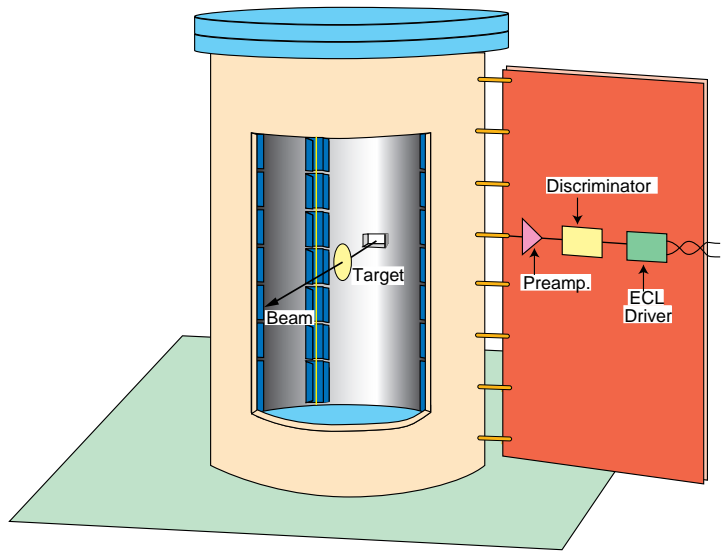,width=9.5cm}
  \caption{Centrality detector}
  \label{fig:centrality}
\end{figure}

\vspace{3mm}
\label{sec:grey}
A graphite target of 0.7~cm length and 0.6~cm diameter with a density of
1.83~g/cm$^3$ has been used. This corresponds to an interaction length
of 1.5\%. The target is housed inside a grey proton detector \cite{bib:nim} 
which has been developed for the control of impact parameter (centrality) 
in p+A collisions and is presented schematically in Fig.~\ref{fig:centrality}.

This detector is a cylindrical proportional counter of 12 cm diameter 
which surrounds the target and has a window in the forward hemisphere
corresponding to the acceptance of the spectrometer inside polar angles
of $<$~45$^o$. Its surface is subdivided into 256 pads which provides 
ample granularity for the counting of the typically less than 8 grey
protons measured per event in light ion applications. A thin
(200~$\mu$m) copper sheet on the inner surface absorbs nuclear
fragments by range, and an electronics threshold placed at 1.5 times
the minimum ionization deposit cuts high momentum particles as the
grey protons are placed high on the $1/\beta^2$ branch of the 
Bethe-Bloch energy-loss distribution. Grey protons in the momentum
range 0.15 to 1.2~GeV/c reconstructed and identified inside the
spectrometer acceptance are added to the number measured in the
centrality detector. 

\begin{figure}[b!]
  \centering
  \epsfig{file=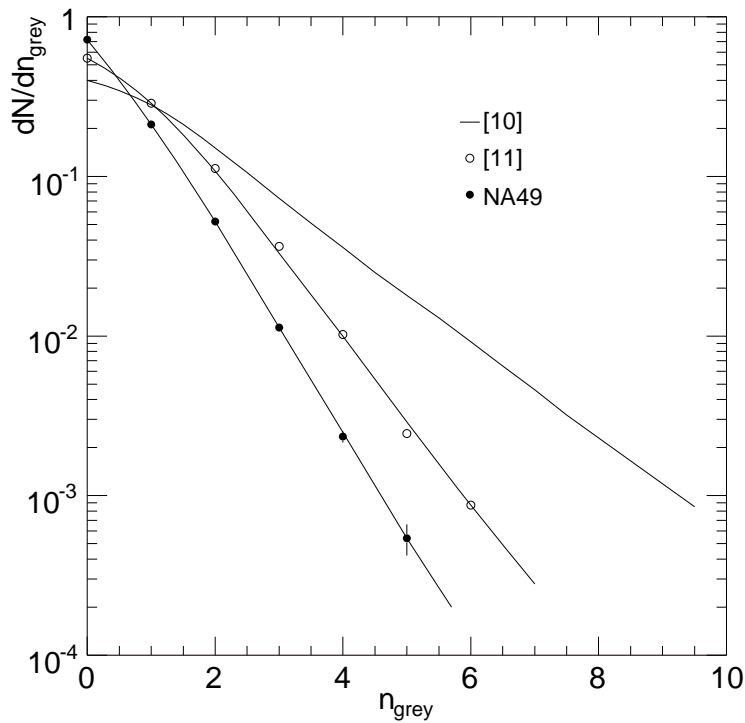,width=10cm}
  \caption{Number of grey protons $n_{\textrm{grey}}$ distribution. 
           The lines through the data points are drawn to guide the eye }
  \label{fig:ncd}
\end{figure}

The resulting number distribution is shown in Fig.~\ref{fig:ncd}, 
together with the prediction by Hegab and H\"ufner 
\cite{bib:hh1} for the total grey proton yield and the measurement 
by Braune et al. \cite{bib:braune} at 50 and 100~GeV/c beam momentum at 
the SPS using a detector of 60\% geometrical acceptance. Compared to this 
the NA49 data correspond to a 45\% effective acceptance which is in part  
due to the steep angular distribution of slow protons in p+C collisions 
\cite{bib:braune} and to the large opening of $\pm$45$^o$
imposed by the spectrometer acceptance. Only about 30\% of protons in this 
angular range with momenta below 1~GeV/c are in fact reconstructed and 
identified by the NA49 tracking system.

Due to the short available data taking period and to the sharp
drop of the event yield as a function of the number of grey protons
(less than 30\% of all events have a measured grey proton), the on-line 
triggering capability on grey protons available in the NA49 trigger system 
could not be used. All data have therefore been obtained in ``minimum bias''
condition without imposing centrality selection. It is nevertheless
possible to use the grey proton information in a sample of $p_T$ integrated
inclusive cross sections in order to study the evolution with centrality
as described in Sect.~\ref{sec:ncd}.
                                                                                
As the range of grey protons at their most probable momentum of 0.3~GeV/c
is only 1~cm in Carbon, the target diameter of 6~mm has been kept at a
minimum with respect to the transverse beam profile \cite{bib:pp_paper} 
in order to reduce the absorption of grey protons by energy loss in the 
target material. The small fraction of beam particles in the tail of the
transverse profile beyond the target radius ($\approx 3\%$) has been 
corrected for in the determination of the trigger cross section.
                                                                                
%
%
\subsection{Trigger cross section}
\vspace{3mm}
\label{sec:trig}
The trigger scheme was the same as in p+p interactions, using an
interaction trigger defined by a small scintillation counter 380~cm
downstream of the target in anti-coincidence with the beam. Due to
the reduction of forward protons in p+A interactions by baryon
number transfer towards central rapidities and due to the corresponding
yield decrease of produced particles at large $x_F$, the systematic 
effects connected to this trigger method are smaller than in p+p
collisions and amount to a reduction of the trigger cross section 
of 9\%$\pm$2\% with respect to the total inelastic cross section. A break 
down of these losses determined by using measured inclusive distributions 
of protons, pions and kaons \cite{bib:unpub} in a Monte Carlo 
calculation, is given in Table~\ref{tab:trigger}.

\begin{table}[h]
  \begin{center}
    \begin{tabular}{|l|r@{.}l|}
      \hline
      $\sigma_{\textrm{trig}}$                    & 210&1 mb      \\ 
      loss from p                                 &  17&1 mb      \\ 
      loss from $\pi$ and K                       &   2&4 mb      \\ 
      contribution from $\sigma_{\textrm{el}}$    &  -3&3 mb      \\
      predicted $\sigma_{\textrm{inel}}$          & 226&3 mb      \\ \hline
      literature value                            & 225&8 mb      \\ \hline
    \end{tabular}
  \end{center}
  \vspace{-2mm}
  \caption{Contributions derived from the detailed Monte Carlo calculation to
           the determination of the measured inelastic cross section 
	   $\sigma_{\textrm{inel}}$}
  \label{tab:trigger}
\end{table}

\begin{figure}[t]
  \centering
  \epsfig{file=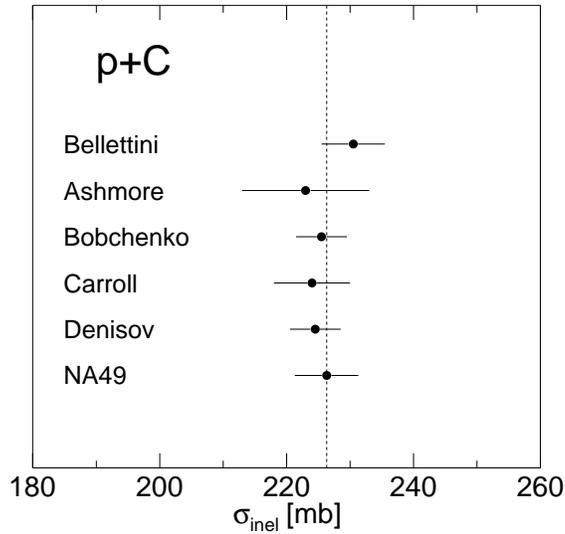,width=7.5cm}
  \caption{Inelastic cross section compared to previous measurements 
           \cite{bib:pc_inel} }
  \label{fig:inel}
\end{figure}

The resulting measured value of the inelastic p+C cross section is 
226.3~mb with an estimated systematic error of 2.5\% (see Table~\ref{tab:sys}). 
It compares well with a compilation of preceding measurements
\cite{bib:pc_inel} giving 225.8~mb as shown in Fig.~\ref{fig:inel}.

The relatively large systematic deviations visible in the reference values,
which have all been obtained in transmission experiments using yield
extrapolation to zero momentum transfer, is noteworthy. It is partially
attributable to the definition of the elastic component and to the
corresponding uncertainty in the slopes of the $t$-distributions used.
Compared to the precision available in the elementary hadron-nucleon
inelastic cross sections, the absolute normalization of inclusive
yields therefore suffers from larger systematic uncertainties.

%
%
\subsection{Event sample and cuts}
\vspace{3mm}
\label{sec:events}

\begin{figure}[b!]
  \centering
  \epsfig{file=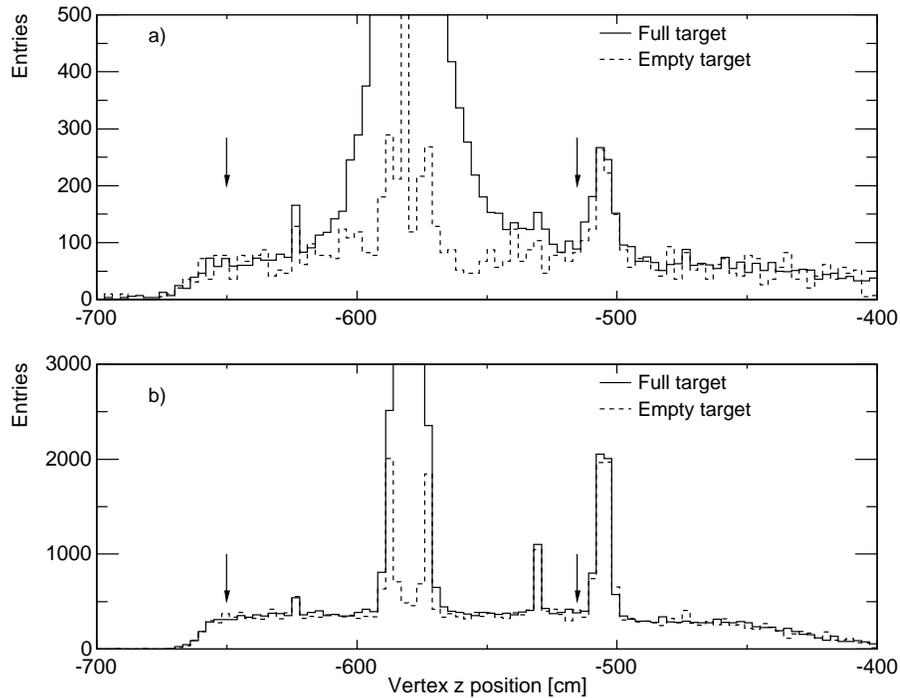,width=12cm}
  \caption{Normalized vertex distributions from full and empty target events with
           selected track multiplicity a) one and b) five and more. The arrows
           indicate the fiducial region }
  \label{fig:zvertex}
\end{figure}

In addition to the cuts imposed on the transverse beam definition as
obtained from the Beam Position Detectors \cite{bib:nim} an additional cut 
on the transverse beam radius at less than 3~mm is imposed due to the
target dimension (Sect.~\ref{sec:grey}). The longitudinal vertex position is
constrained to a fiducial region, as exemplified for two values of 
charged multiplicity in Fig.~\ref{fig:zvertex}, where the distribution 
of detector material in the vicinity of the target is clearly visible. 

The combined cuts result in the event sample given in Table~\ref{tab:events}
where in particular the reduction of the empty target fraction from 30\% 
in the total sample to 16\% after cuts is noteworthy. Due to the small correction
imposed by the empty target contribution (see Sect.~\ref{sec:empty}), the fraction
of running time spent on empty target could be kept at about 5\% only.

\begin{table}[h]
  \begin{center}
    \begin{tabular}{|cc|cc|}
      \hline
      \multicolumn{2}{|c|}{Events taken} &
      \multicolumn{2}{|c|}{Events after selection} \\
      Full target & Empty target & Full target & Empty target \\
      \hline
      535.7 k     & 31.2 k       &  377.6 k    & 11.8 k     \\
      \hline 
    \end{tabular}
  \end{center}
  \vspace{-2mm}
  \caption{Data sample}
  \label{tab:events}
\end{table}
 
%
%
\subsection{Acceptance coverage, binning and statistical errors}
\vspace{3mm}
\label{sec:bin_scheme}

\begin{figure}[b]
  \centering
  \epsfig{file=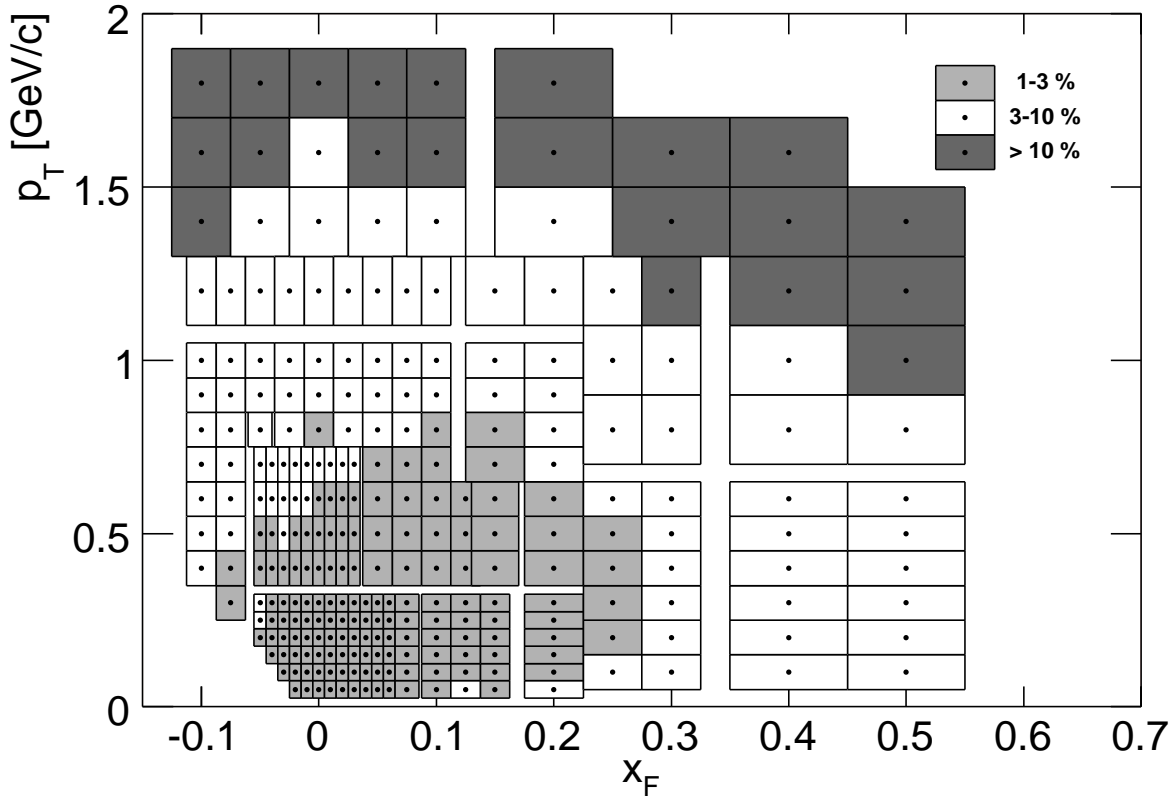,width=16cm}
  \caption{Binning scheme in $x_F$ and $p_T$ together with information of the 
           statistical error }
  \label{fig:bin_scheme}
\end{figure}

Due to the forward-backward asymmetry of p+A interactions a substantial
effort was spent in exploiting the available acceptance in the region
of negative $x_F$. Here limitation are imposed both by the detector
acceptance itself and by the particle identification problems in the 
region near minimum ionization. The resulting phase space coverage
with 270 bins per charge is presented in Fig.~\ref{fig:bin_scheme} where 
also the statistical errors per bin are indicated. 

The reduced statistics compared to the much larger event sample in 
p+p collisions \cite{bib:pp_paper} results in larger bins generally
and a limitation to the regions $p_T\leq$~1.8~GeV/c and $x_F\leq$~0.5. 
It nevertheless provides an unprecedented overall coverage which allows 
a detailed study of the important cross-over region between target 
and projectile fragmentation for the first time.

%
%
\subsection{Particle identification}
\begin{figure}[h]
  \centering
  \epsfig{file=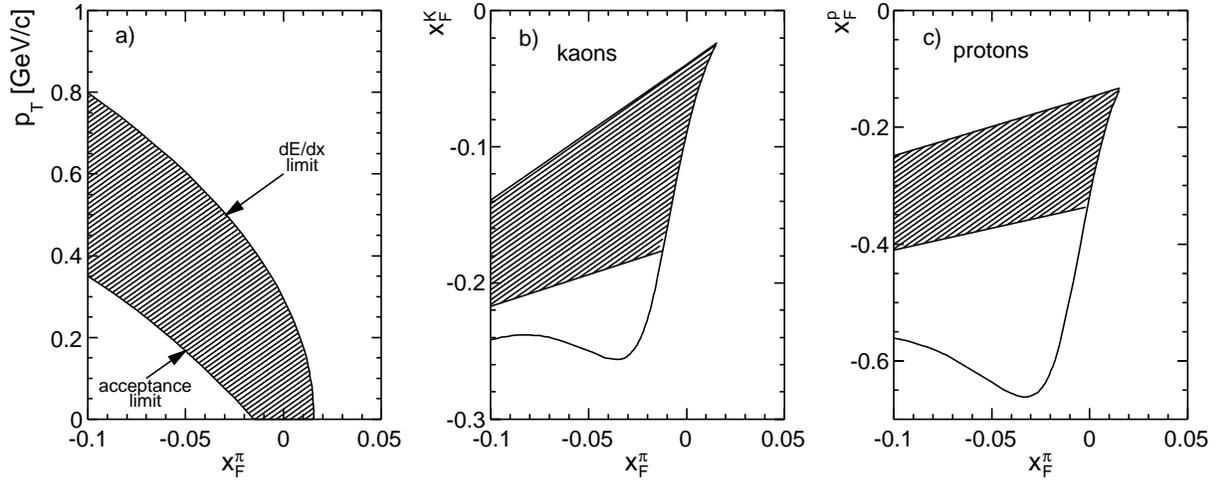,width=16cm}
  \caption{$dE/dx$ cross-over region: a) in ($x_F$,$p_T$) plane of the pions,
           b) in ($x_F^\pi$,$x_F^K$) plane, and c) in ($x_F^\pi$,$x_F^p$) plane.
           The unhatched areas correspond to the regions where pions can
           be identified due to the 1/$\beta^2$ increase of energy loss of the kaons
	   and protons, respectively }
  \label{fig:back_kin}
\end{figure}

\vspace{3mm}
The procedures for the extraction of pion yields from the energy loss
distributions measured in each bin are identical to the ones used
in p+p collisions \cite{bib:pp_paper} for most of the forward hemisphere. 
The extended coverage of the backward hemisphere together with the
forward-backward asymmetry of p+C interactions necessitates
an extension of the methods developed in \cite{bib:pp_paper} for the 
treatment of the lab momentum region below 3~GeV/c where the energy 
loss functions of pions, kaons and protons approach each other. 
The kinematic situation in this region is indicated in Fig.~\ref{fig:back_kin}.

\begin{figure}[b]
  \centering
  \epsfig{file=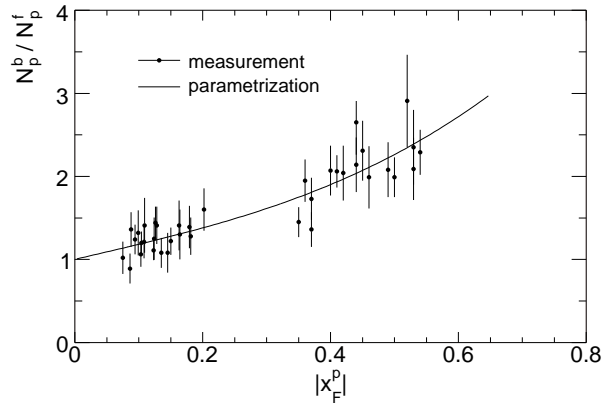,width=8cm}
  \caption{Backward/forward yield ratio for protons}
  \label{fig:fb_proton}
\end{figure}

Using the pion mass in the transformation from $x_F$ to lab momentum
the critical zone is defined by the hatched area in Fig.~\ref{fig:back_kin}a
where the upper limit corresponds to $p_{\textrm{lab}}$~= 3~GeV/c 
and the lower limit
traces the $p_T$ cut-off used in the data extraction (Fig.~\ref{fig:bin_scheme}). 
This area is mapped into the $x_F$ regions for kaons (Fig.~\ref{fig:back_kin}b) 
and protons (Fig.~\ref{fig:back_kin}c) as a function of 
$x_F^{\textrm{pion}}$ 
when using the proper masses in the corresponding Lorentz transformations. 

For protons the lower part of the critical zone is again available
for $dE/dx$ extraction due to the rapid increase of their energy
loss in the $1/\beta^2$ region of the Bethe-Bloch function. The 
remaining band at -0.4~$<x_F^{\textrm{proton}}<$~-0.2 
is treated by reflecting
the $x_F$ bins into the forward hemisphere and by using an interpolation
of the measured backward/forward yield ratios shown in 
Fig.~\ref{fig:fb_proton}.

\begin{figure}[t]
  \centering
  \epsfig{file=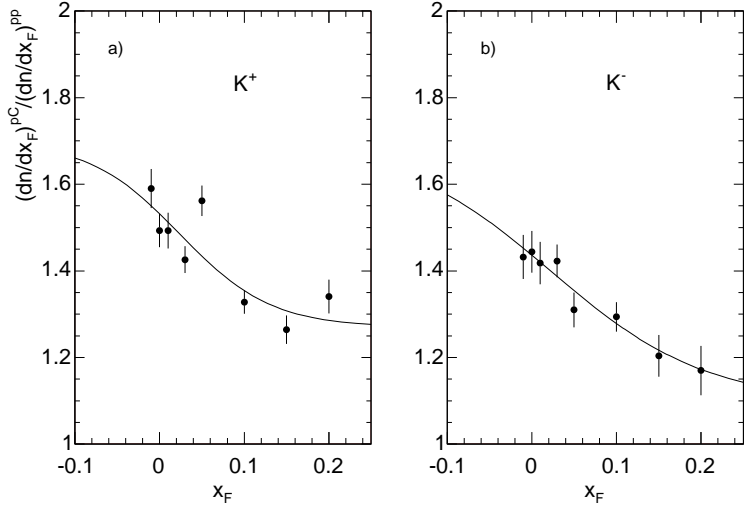,width=10cm}
  \caption{$(dn/dx_F)^{\textrm{pC}}$/$(dn/dx_F)^{\textrm{pp}}$ 
           ratio as a function of $x_F$ of the kaons for a) K$^+$ and b) K$^-$}
  \label{fig:kaon}
\end{figure}

The $\overline{\mbox{p}}/\pi^-$ and K/$\pi$ ratios are smaller than 
10\% for all bins in the critical area. Here again the method of bin 
reflection was used by imposing an extrapolation of the measured yield 
ratio with respect to p+p interactions into the backward hemisphere as 
shown in Fig.~\ref{fig:kaon} for kaons. In this case the ratio in the far 
backward region corresponds to the mean number of projectile collisions 
predicted from the inelastic p+C and p+p cross sections \cite{bib:hh2}. 
For kaons the different loss rates from weak decays in the forward and 
backward hemispheres were taken into account in the yield determination.

%
%
\section{Evaluation of invariant cross sections and corrections}
\vspace{3mm}
\label{sec:corr}
The invariant inclusive cross section

\begin{equation}
  f(x_F,p_T) = E(x_F,p_T) \cdot \frac{d^3\sigma}{dp^3} (x_F,p_T)
\end{equation}
is experimentally defined by the measured quantity \cite{bib:pp_paper}

\begin{equation}
  f_{\textrm{meas}}(x_F,p_T,\Delta p^3) =
  E(x_F,p_T,\Delta p^3) \cdot \frac{\sigma_{\textrm{trig}}}
  {N_{\textrm{ev}}} \cdot
  \frac{\Delta n(x_F,p_T,\Delta p^3)}{\Delta p^3}~~~,
\end{equation}
where $\Delta p^3$ is the finite phase space element defined by the
bin width.

As in \cite{bib:pp_paper} several steps of normalization and correction 
have to be applied in order to make $f_{\textrm{meas}}(x_F,p_T,\Delta p^3)$ 
approach $f(x_F,p_T)$.
The determination of the trigger cross section and its deviation
from the total inelastic cross section has been discussed above.
The corrections for pion weak decay and absorption in the detector
material are identical to p+p and will not be discussed here.
The remaining corrections which are numerically different in p+C
interactions are:

\vspace{2mm}
\begin{itemize}
\item treatment of empty target contribution
\item re-interaction in the target volume
\item effect of the interaction trigger
\item feed-down from weak decays of strange particles
\item effect of finite bin width.
\end{itemize}
\vspace{2mm}

These corrections will be described and quantified below.

%
%
\subsection{Empty target contribution}
\begin{figure}[h]
  \centering
  \epsfig{file=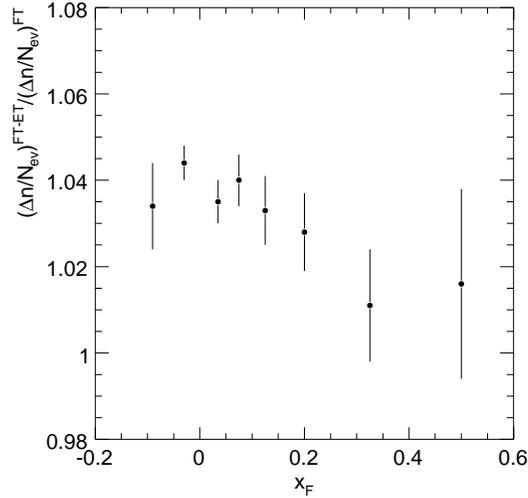,width=7cm}
  \caption{Correction factor applied to account for empty target contribution as a
           function of $x_F$ for the average of $\pi^+$ and $\pi^-$ }
  \label{fig:empty}
\end{figure}

\vspace{3mm}
\label{sec:empty}
In the determination of the normalized quantity 

\begin{equation}
  \left(\frac{\Delta n}{N_{\textrm{ev}}}\right)^{\textrm{FT-ET}} = \frac{1}{1-\epsilon}
  \left(\left(\frac{\Delta n}{N_{\textrm{ev}}}\right)^{\textrm{FT}} -
  \epsilon \left(\frac{\Delta n}{N_{\textrm{ev}}}\right)^{\textrm{ET}} \right) ,
  \label{eq:empty}
\end{equation}
the ratio $\epsilon$ of empty over full trigger rates has been reduced from
0.3 to 0.16 by the cuts described in Sect.~\ref{sec:events} above. 
In addition, the bulk of the remaining empty target yield is produced in Mylar
foils and air (Fig.~\ref{fig:zvertex}). These materials are sufficiently 
close to Carbon to make the normalized bin contents 
$(\Delta n/N_{\textrm{ev}})^{\textrm{FT}}$ and 
$(\Delta n/N_{\textrm{ev}})^{\textrm{ET}}$ 
approximately equal. The deviation of the complete
normalized yield from the full target yield alone is therefore
expected to be small and to be essentially defined by the different
fraction of empty events in full and empty target conditions.
This is demonstrated in Fig.~\ref{fig:empty} where the ratio between the
bin contents for full-empty target and full target alone is shown
as a function of $x_F$.

Within the limits of the experimental accuracy no differences between
$\pi^+$ and $\pi^-$ and no dependencies on $p_T$ have been observed.  

%
%
\subsection{Target re-interaction}
\vspace{3mm}
The Carbon target has an interaction length
of 1.5\% which corresponds to only 55\% of the length of the hydrogen
target used in p+p interactions. The expected re-interaction correction
is therefore below the 2\% level even in the extreme backwards bins and has
been scaled down accordingly from the values obtained in \cite{bib:pp_paper}.

%
%
\subsection{Trigger bias correction}
\vspace{3mm}
Several effects contribute to a modification of the correction for the
trigger bias which is introduced by the interaction trigger in p+C
as compared to p+p collisions. They all lead to a reduction of the
correction to the p+C data.

\vspace{2mm}
\begin{itemize}
\item Due to baryon number transfer towards central rapidity (``stopping'')
      there are less forward protons hence a smaller probability to
      veto events by the trigger counter.
\item Due to the correlated steepening of the $x_F$ distributions of produced
      particles (see Sect.~\ref{sec:res}) there is again a reduction of 
      the veto probability. These two effects combine to explain the 
      higher inelastic trigger efficiency of 93\% in p+C as compared to 
      89\% in p+p collisions.
\item Unlike in p+p events, there is a long-range correlation between target
      fragmentation and forward particle density. Large $x_F$ protons are
      correlated with single projectile collisions yielding small backward  
      multiplicities, whereas multiple projectile collisions result in high
      target multiplicity and low forward yields. This correlation reduces
      the effect of the interaction trigger in the backward hemisphere.
\end{itemize}
\vspace{2mm}

\begin{figure}[h]
  \centering
  \epsfig{file=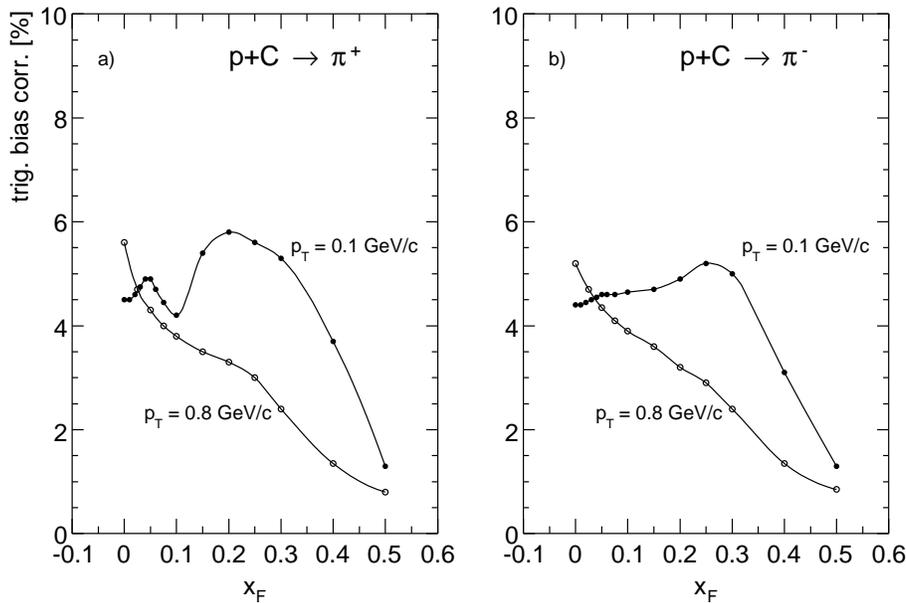,width=12cm}
  \caption{Trigger bias correction as a function of $x_F$ at various $p_T$ for 
           a) $\pi^+$ and b) $\pi^-$ }
  \label{fig:tr_bias}
\end{figure}

The trigger bias correction has been obtained using the method developed
in \cite{bib:pp_paper} by artificially increasing the diameter of the trigger 
counter in the analysis and extrapolating to surface zero. The resulting 
corrections are shown in Fig.~\ref{fig:tr_bias} as a function of $x_F$ 
for two values of $p_T$. The correction is smaller than in the p+p case
\cite{bib:pp_paper} and thus confirms the reduction quoted above.


%
%
\subsection{Feed-down correction}
\vspace{3mm}
A principle problem in the determination of the feed-down correction 
induced by the weak decays of strange particles lies in the absence
of data on K$^0_S$ and strange baryon production in light ion collisions.
The corresponding yields have therefore been determined from the NA49
data directly using the following yield ratios with respect to p+p
interactions:

\begin{figure}[t]
  \centering
  \epsfig{file=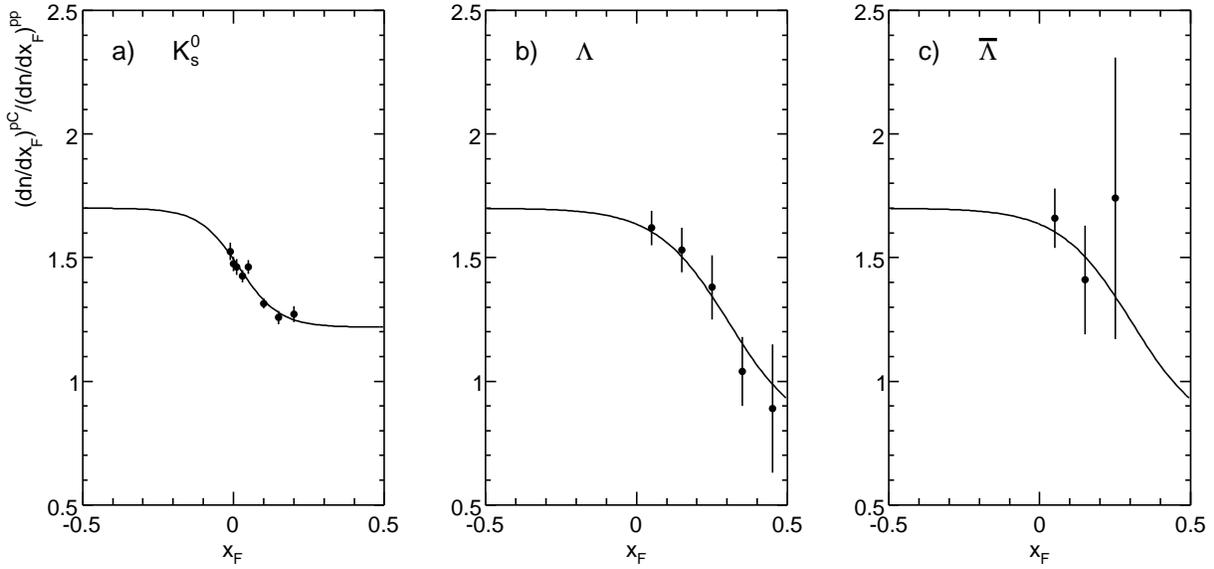,width=16cm}
  \caption{Yield ratios for  a) $K^0_S$, b) $\Lambda$ and c) $\overline{\Lambda}$}
  \label{fig:s_ratio}
\end{figure}

\begin{figure}[b]
  \centering
  \epsfig{file=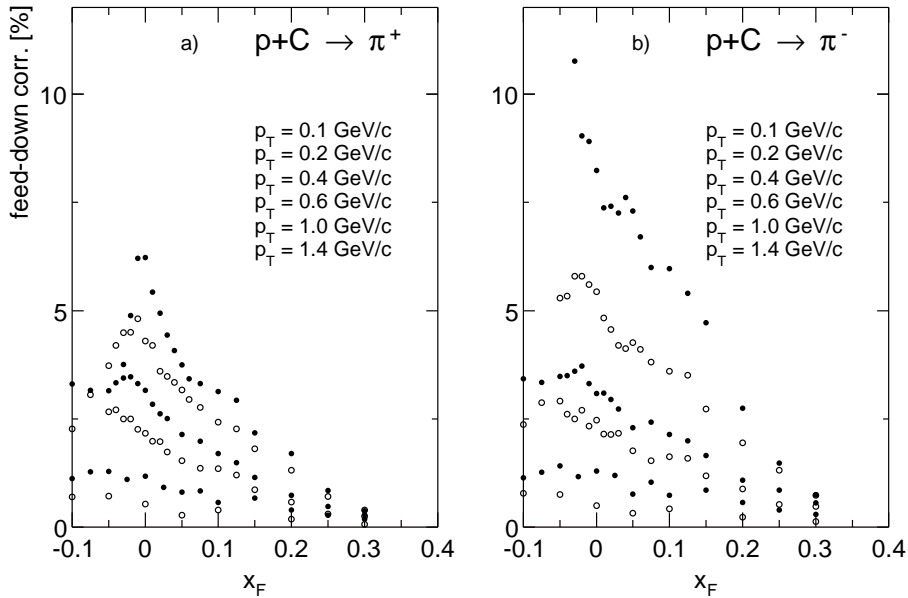,width=12cm}
  \caption{Feed-down correction to pion from weak decays for a) $\pi^+$ and 
           b) $\pi^-$ }
  \label{fig:feeddown}
\end{figure}

\vspace{2mm}
\begin{itemize}
\item The K$^0_S$ yield is extracted from (K$^+$ + K$^-$)/2. The measured ratio is 
      extrapolated into the backward hemisphere using a two-component 
      superposition picture \cite{bib:hgf} as shown in Fig.~\ref{fig:s_ratio}a. 
      As the kaon yields do not suffer from isospin effects \cite{bib:hgf} this 
      extrapolation is straight-forward concerning the target contribution.
\item The evolution of $\Lambda$ and $\bar{\Lambda}$ yields relative to 
      p+p is obtained from p+$\pi^-$ and $\overline{\mbox{p}}$+$\pi^+$ 
      mass distributions exploiting the event mixing technique described in 
      \cite{bib:drijard} and using vertex tracks both for the baryon and for 
      the meson involved. The resulting yield ratios are shown in 
      Figs.~\ref{fig:s_ratio}b and c. For the extrapolation into the 
      backward hemisphere the two-component superposition picture is 
      again used. No isospin effects are present in this extrapolation.
\end{itemize}
\vspace{2mm}

The yield ratios for $\Sigma^{\pm}$ are derived from the $\Lambda$ 
parametrization using the $\Sigma/\Lambda$ ratios from p+p in 
the projectile hemisphere. In the target fragmentation region the expected 
isospin effects are taken into account. 

The resulting overall feed-down corrections are shown in 
Fig.~\ref{fig:feeddown} as a function of $x_F$ for several values of $p_T$.

%
%
\subsection{Binning correction}
\begin{figure}[h]
  \centering
  \epsfig{file=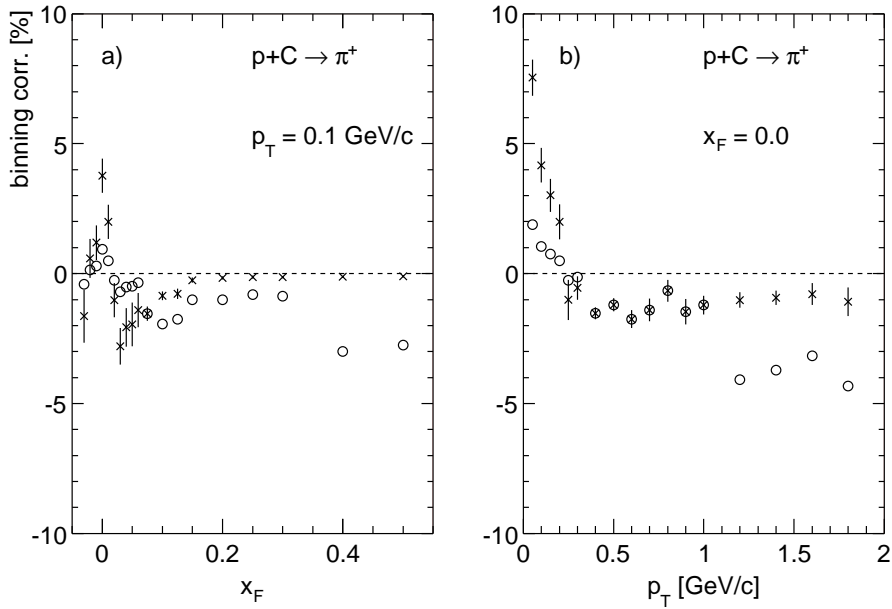,width=12cm}
  \caption{Correction due to the binning in a) $x_F$ and b) $p_T$.  The crosses
           represent the correction at fixed bin widths of 
           $\Delta x_F$~= 0.02 and $\Delta p_T$~= 0.1~GeV/c, respectively and the open
           circles describe the correction for the bins actually used }
  \label{fig:binning}
\end{figure}

\vspace{3mm}
The correction for finite bin width follows the scheme developed in 
\cite{bib:pp_paper} determining the deviation of the real cross section at
the bin center from the measured one (averaged over the bin) using the local 
second derivative of the 
particle density distribution. The derivative is calculated using the 
experimental data and therefore does not depend on specific parametrizations. 
Typical values of the resulting correction are shown in Fig.~\ref{fig:binning}
for low $p_T$ as a function of $x_F$ and for $x_F$~= 0 as a function of 
$p_T$, both for constant bin widths of $\Delta x_F$~= 0.02 and 
$\Delta p_T$~= 0.1~GeV/c, and for the bin widths actually used in the 
data extraction.

Due to the increased average bin width in p+C the values are somewhat
larger than in p+p, but typically stay below the $\pm$2\% limit, except
for $x_F\geq$~0.4 and $p_T\geq$~1.2 GeV/c.

%
%
\subsection{Systematic errors}
\vspace{3mm}
An estimation of the systematic errors induced by the overall 
normalization and by the applied corrections is given in Table~\ref{tab:sys}.

\begin{table}[h]
  \begin{center}
    \begin{tabular}{|l|r|}
      \hline
      Normalization                                 &  2.5\%   \\ \hline
      Tracking efficiency                           &  0.5\%   \\ \hline
      Trigger bias                                  &    1\%   \\ \hline
      Feed-down                                     &  1-2.5\% \\ \hline
      Detector absorption                           &          \\
      Pion decay $\pi\rightarrow\mu + \nu_{\mu}$    &  0.5\%   \\
      Re-interaction in the target                  &          \\ \hline
      Binning                                       &  0.5\%   \\ \hline \hline
      Total(upper limit)                            &  7.5\%   \\ \hline
      Total(quadratic sum)                          &  3.8\%   \\ \hline
    \end{tabular}
  \end{center}
  \vspace{-2mm}
  \caption{Summary of systematic errors}
  \label{tab:sys}
\end{table}

\begin{figure}[h]
  \centering
  \epsfig{file=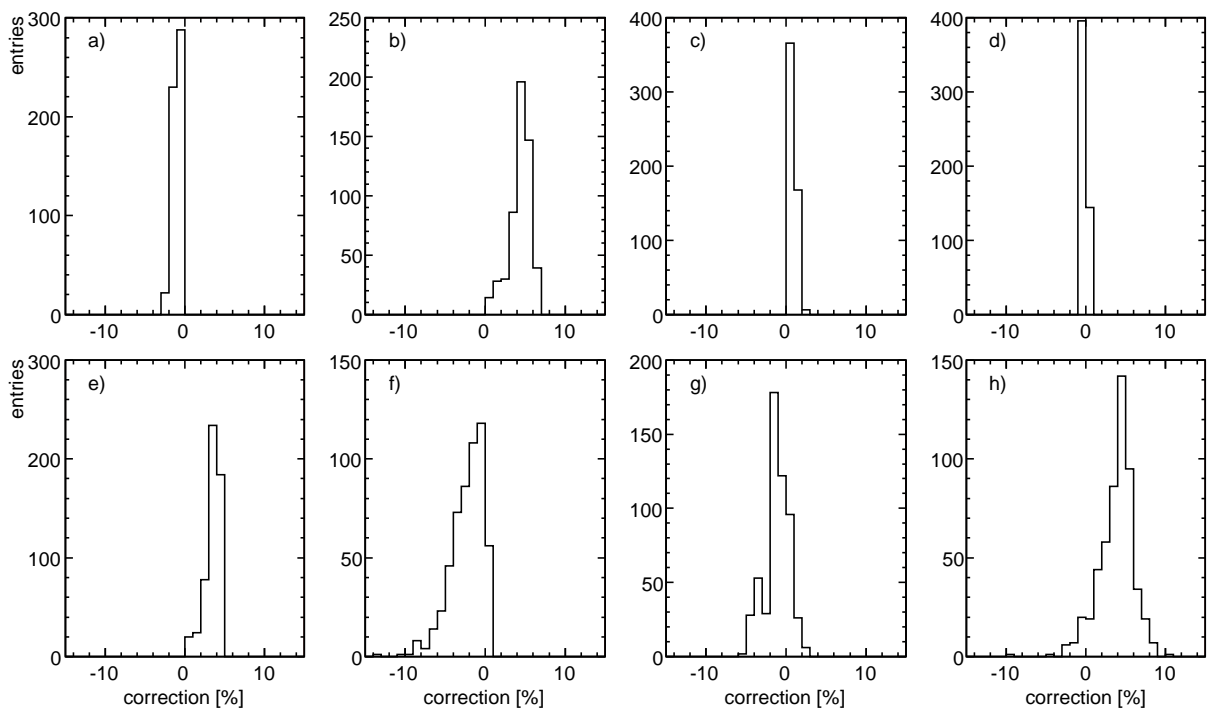,width=16cm}
  \caption{Distribution of correction for a) target re-interaction, 
           b) trigger bias, c) absorption in detector material, 
	   d) pion decay, e) empty target contribution, f) feed-down, 
	   g) binning, and h) resulting total correction}
  \label{fig:sys}
\end{figure}

An upper limit of 7.5\% from linear addition of the error sources
and an rms error of 3.8\% from quadratic summation are obtained.
Further information on the bin-by-bin variation of the different
error contributions is contained in Fig.~\ref{fig:sys} which demonstrates that
the single errors fluctuate around well-defined mean values with
limited skewness and tails. 

%
%
\section{Results}
\vspace{3mm}
\label{sec:res}
The set of double differential invariant cross sections obtained from
the data analysis and correction procedures described above forms,
by its dense coverage of the available phase space in 270 bins per
charge, an internally consistent ensemble that reveals, as in
the case of p+p interactions, complex structures which pervade
both transverse and longitudinal momentum dependencies. These structures
defy simple parametrization with straight-forward arithmetic
expressions. To make full use of the consistency of the 
data set and to allow for optimum precision in the determination
of integrated cross sections, a numerical interpolation scheme has
therefore been used that relies on local continuity in both kinematic
variables and allows for limited extrapolation into the inaccessible
regions of phase space. This chapter summarizes the numerical
information in data tables, gives a set of distributions as a function
of $p_T$, $x_F$ and rapidity $y$, and shows the comparison to existing data.

%
%
\subsection{Data tables, distributions and interpolation scheme}
\vspace{3mm}
Tables~\ref{tab:pip} and \ref{tab:pim} present the invariant inclusive 
cross sections for $\pi^+$ and $\pi^-$ respectively. They correspond 
to the binning scheme discussed in Sect.~\ref{sec:bin_scheme} above 
and reflect the attempt to cover the kinematic 
plane as completely as possible given the limited statistical accuracy  
of the data sample.

\begin{figure}[h]
  \centering
  \epsfig{file=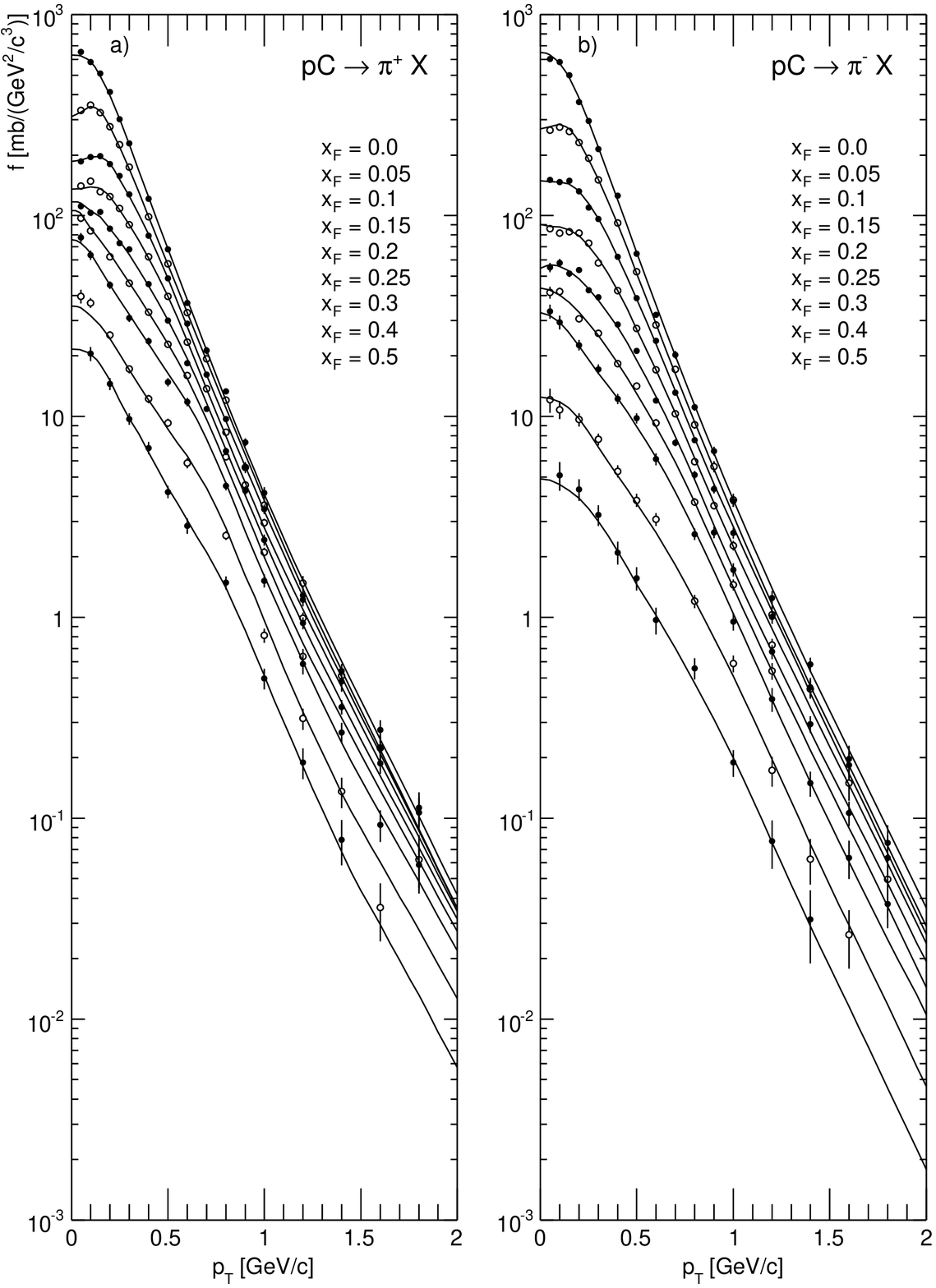,width=16cm}
  \caption{Invariant cross section as a function of $p_T$ at fixed $x_F$ for 
           a) $\pi^+$ and b) $\pi^-$ produced in p+C collisions at 158~GeV/c}
  \label{fig:pt_dist}
\end{figure}

\begin{figure}[h]
  \centering
  \epsfig{file=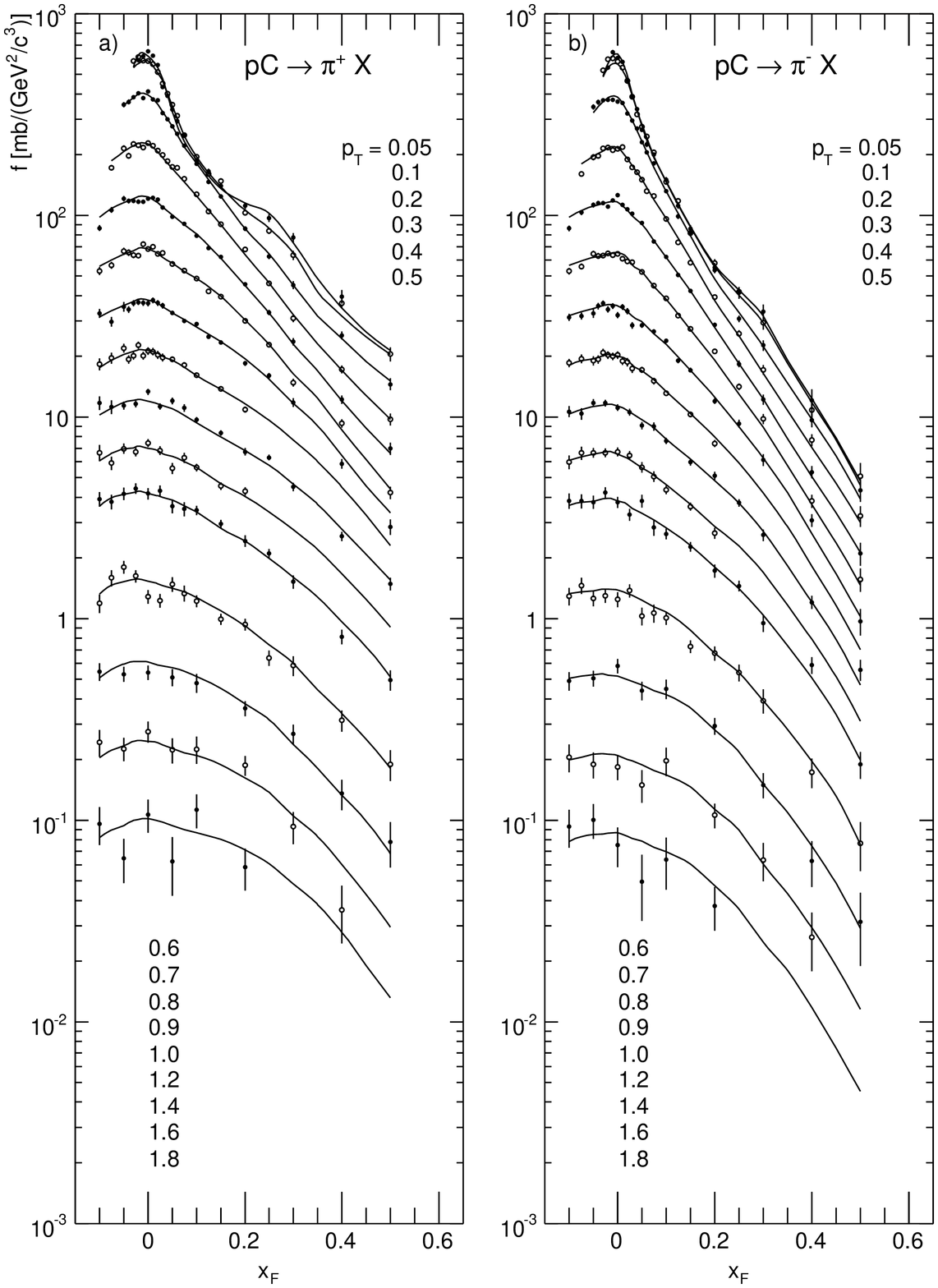,width=16cm}
  \caption{Invariant cross section as a function of $x_F$ at fixed $p_T$ for 
           a) $\pi^+$ and b) $\pi^-$ produced in p+C collisions at 158~GeV/c}
  \label{fig:xf_dist}
\end{figure}

The distributions of the data in $p_T$ and $x_F$ are
shown in Figs.~\ref{fig:pt_dist} and \ref{fig:xf_dist} 
respectively. Here the full lines represent the data interpolation 
mentioned above.

\begin{figure}[h]
  \centering
  \epsfig{file=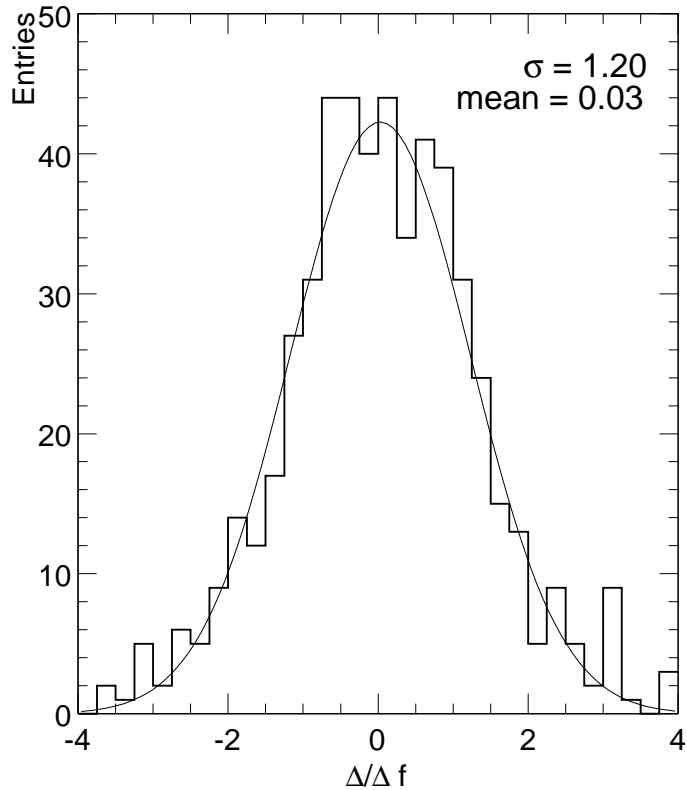,width=9.5cm}
  \caption{Histogram of the differences $\Delta$ between the measured invariant
           cross sections and the corresponding interpolated values ($\pi^+$ 
	   and $\pi^-$ combined) divided by the experimental uncertainty $\Delta f$ 
	   of the data points}
  \label{fig:norm}
\end{figure}

In order to demonstrate the statistical consistency of the 
interpolation scheme, the distribution of the differences between
data and interpolation, normalized with the statistical error
in each data point, is shown in Fig.~\ref{fig:norm}. The shape and 
width of the distribution comply with the expected Gaussian behaviour.

\begin{table}
\scriptsize
\begin{center}
\begin{tabular}{|c|c@{\hspace{2mm}}c@{\hspace{0.8mm}}
                  |c@{\hspace{2mm}}c@{\hspace{0.8mm}}
		  |c@{\hspace{2mm}}c@{\hspace{0.8mm}}
                  |c@{\hspace{2mm}}c@{\hspace{0.8mm}}
	          |c@{\hspace{2mm}}c@{\hspace{0.8mm}}
	          |c@{\hspace{2mm}}c@{\hspace{0.8mm}}
	          |c@{\hspace{2mm}}c@{\hspace{0.8mm}}
	          |c@{\hspace{2mm}}c@{\hspace{0.8mm}}
	          |c@{\hspace{2mm}}c@{\hspace{0.8mm}}|}
\hline
\multicolumn{19}{|c|}{$f(x_F,p_T), \Delta f$} \\
\hline
$p_T \backslash x_F$ & \multicolumn{2}{|c|}{-0.1} & \multicolumn{2}{|c|}{-0.075}  
                     & \multicolumn{2}{|c|}{-0.05} & \multicolumn{2}{|c|}{-0.04}  
                     & \multicolumn{2}{|c|}{-0.03} & \multicolumn{2}{|c|}{-0.025}
                     & \multicolumn{2}{|c|}{-0.02} & \multicolumn{2}{|c|}{-0.01}  
                     & \multicolumn{2}{|c|}{0.0} \\ \hline
0.05 && && && && && && &589.8&2.41 &616.9&2.50 &652.3&1.59\\ \hline
0.1  && && && && &584.0&2.62 && &611.0&1.77 &587.3&1.59 &581.7&1.68\\\hline 
0.15 && && && &445.2&2.80 &488.4&1.89 && &509.8&1.81 &502.0&1.66 &508.2&1.55\\ \hline
0.2 && && &353.4&2.99 &365.7&2.70 &385.2&1.93 && &403.2&1.85 &382.1&1.77
    &411.9&1.65\\ \hline
0.25 && &&  &268.6&3.20 &284.5&2.55 &308.2&2.28 && &287.1&2.14 &300.3&1.97
     &302.2&1.89\\ \hline
0.3 && &172.2&2.43 &214.8&3.82 &197.6&3.34 &225.8&2.64 && &222.9&2.37
    &216.7&2.23 &228.4&2.12\\ \hline
0.4 &86.5&3.32 &106.4&3.33 &121.2&3.10 &117.8&2.80 &118.1&2.41 &&
    &117.0&2.24 &116.9&2.24 &121.0&1.81\\ \hline
0.5 &53.0&4.63 &56.6&4.00 &66.6&4.06 &65.4&3.69 &63.6&3.41 &&
    &63.0&2.57 &71.9&2.38 &68.0&2.43\\ \hline
0.6 & 32.7&5.14 & 29.7&5.46 & 35.3&5.54 & 34.1&3.66 & 36.7&3.41 && & 37.1&3.31
    & 36.8&3.28 & 36.6&3.32\\ \hline
0.7 & 18.3&6.82 & 19.6&6.43 & 21.9&5.03 &19.30&4.78 &20.13&4.59 && &22.73&4.26
    &20.11&4.49 &21.28&4.28\\ \hline
0.8 &11.76&7.85 &11.22&7.91 &11.39&4.46 && && &11.61&3.82 && && &13.35&3.53\\ \hline
0.9 & 6.65&9.43 & 5.91&7.49 & 6.97&5.71 && && & 6.72&4.95 && && & 7.43&4.52\\ \hline
1.0 & 3.92&8.52 & 3.79&8.19 & 4.16&6.91 && && & 4.43&6.03 && && & 4.17&6.42\\ \hline
1.2 & 1.19&10.5 & 1.59&9.19 & 1.80&7.44 && && & 1.62&6.91 && && & 1.285&7.73\\
\hline
1.4 & 0.545&10.0 && & 0.530&8.5 && && && && && & 0.541&8.47\\ \hline
1.6 & 0.244&15.3 && & 0.226&13.3 && && && && && & 0.275&12.0\\ \hline
1.8 & 0.096&21.3 && & 0.065&24.7 && && && && && & 0.107&18.8\\ \hline
\hline
$p_T \backslash x_F$ & \multicolumn{2}{|c|}{0.01} & \multicolumn{2}{|c|}{0.02}
                     & \multicolumn{2}{|c|}{0.025} & \multicolumn{2}{|c|}{0.03}
                     & \multicolumn{2}{|c|}{0.04} & \multicolumn{2}{|c|}{0.05}
                     & \multicolumn{2}{|c|}{0.06} & \multicolumn{2}{|c|}{0.075}
                     & \multicolumn{2}{|c|}{0.1}\\ \hline
0.05 &620.4&1.75 &555.9&2.03 && &432.9&2.10 &390.7&2.01 &334.4&2.41 &293.4&2.81
     &249.6&2.34 &186.3&2.76\\ \hline
0.1 &561.5&1.62 &515.0&1.51 && &451.6&1.69 &401.9&1.74 &353.7&2.04 &311.8&2.39
    &249.7&2.10 &195.6&2.45\\ \hline
0.15 &488.0&1.52 &451.4&1.59 && &407.1&1.87 &360.3&1.51 &324.8&1.76 &287.8&2.07
     &247.8&1.69 &197.1&2.03\\ \hline
0.2 &377.0&1.62 &373.1&1.58 && &322.5&1.89 &300.2&1.74 &276.8&1.66 &254.2&1.91
    &222.0&1.59 &180.0&1.85\\ \hline
0.25 &303.3&1.73 &276.3&1.73 && &263.0&1.90 &237.6&1.83 &224.7&1.94 &211.0&2.19
     &188.8&1.52 &157.5&1.69\\ \hline
0.3 &222.1&1.97 &209.5&1.98 && &199.6&2.07 &185.9&2.34 &174.1&2.13 &172.3&2.29
    &152.1&1.59 &127.5&1.76\\ \hline
0.4 &122.2&1.79 &119.6&1.75 && &109.7&1.90 && & 98.3&1.39 && & 91.0&1.34
    &79.4&1.39\\ \hline
0.5 & 69.8&2.42 & 64.5&2.54 && & 65.2&2.61 && & 57.4&1.89 && & 53.2&2.01
    &48.73&1.62\\ \hline
0.6 & 38.0&3.25 & 36.8&3.34 && & 35.7&3.44 && &32.88&2.39 && &29.98&2.56
    &28.95&1.98\\ \hline
0.7 &21.10&4.28 &20.32&4.43 && &19.74&4.56 && &19.31&3.06 && &18.06&3.40
    &16.10&2.30\\ \hline
0.8 && && &11.28&3.72 && && &12.03&3.84 && &11.09&4.17 & 9.69&3.15\\ \hline
0.9 && && & 6.82&4.89 && && & 5.57&6.32 && & 6.28&5.61 & 5.61&4.30\\ \hline
1.0 && && & 4.30&6.19 && && & 3.61&7.93 && & 3.50&8.18 & 3.45&5.43\\ \hline
1.2 && && & 1.228&7.83 && && & 1.48&8.21 && & 1.33&8.96 & 1.220&6.85\\ \hline
1.4 && && && && && & 0.512&9.62 && && & 0.479&10.8\\ \hline
1.6 && && && && && & 0.223&14.3 && && & 0.225&15.4\\ \hline
1.8 && && && && && & 0.062&32.4 && && & 0.113&19.2\\ \hline
\hline
$p_T \backslash x_F$ & \multicolumn{2}{|c|}{0.125} & \multicolumn{2}{|c|}{0.15}
                     & \multicolumn{2}{|c|}{0.2} & \multicolumn{2}{|c|}{0.25}
                     & \multicolumn{2}{|c|}{0.3} & \multicolumn{2}{|c|}{0.4}
                     & \multicolumn{2}{|c|}{0.5} & \multicolumn{2}{|c|}{}
                     & \multicolumn{2}{|c|}{}\\ \hline
0.05 &165.1&3.39 &140.3&3.11 &111.4&3.59 & 96.9&4.47 & 77.8&5.33 & 39.6&7.63 &&
     && &&\\ \hline
0.1 &160.5&3.00 &147.9&2.69 &103.1&3.26 & 83.7&3.26 & 63.5&5.38 & 36.7&5.56 & 20.5&8.14
    && &&\\ \hline
0.15 &161.6&2.48 &131.2&2.36 &104.3&2.80 && && && && && &&\\ \hline
0.2 &146.8&2.28 &124.3&2.16 & 86.0&2.58 & 62.4&2.76 & 45.1&4.36 & 25.4&4.61 &14.54&6.79
    && &&\\ \hline
0.25 &130.8&2.04 &108.7&1.99 & 72.9&2.52 && && && && && &&\\ \hline
0.3 &104.7&2.16 & 90.2&2.00 & 67.8&2.38 & 46.1&2.79 & 30.9&4.32 &17.19&4.53 & 9.73&6.74
    && &&\\ \hline
0.4 & 68.7&1.64 &62.34&1.50 &45.50&2.06 &33.00&2.84 &23.63&4.21 &12.21&4.61 & 6.98&6.93
    && &&\\ \hline
0.5 &41.95&1.95 &39.51&1.68 &29.98&2.28 &22.82&3.11 &14.82&4.76 & 9.29&4.73 & 4.22&8.48
    && &&\\ \hline
0.6 &24.95&2.29 &23.46&2.35 &18.43&2.62 &16.02&3.29 &11.81&4.84 & 5.86&5.67 & 2.85&8.76
    && &&\\ \hline
0.7 && &13.76&2.55 &10.88&3.44 && && && && && &&\\ \hline
0.8 && & 8.33&3.17 & 6.72&4.28 & 6.30&3.32 & 4.51&4.87 & 2.56&5.18 & 1.49&7.43
    && &&\\ \hline
0.9 && & 4.55&4.07 & 4.28&4.97 && && && && && &&\\ \hline
1.0 && & 2.95&4.87 & 2.43&6.37 & 2.11&5.14 & 1.52&7.57 & 0.812&8.23 & 0.496&11.6  
    && &&\\ \hline
1.2 && & 0.992&6.32 & 0.936&6.68 & 0.639&8.72 & 0.585&11.4 & 0.313&12.0 & 0.190&17.4
    && &&\\ \hline
1.4 && && & 0.358&8.33 && & 0.268&11.2 & 0.136&17.2 & 0.078&25.3 &&
    &&\\ \hline
1.6 && && & 0.187&11.4 && & 0.093&18.0 & 0.036&32.0 && && &&\\ \hline
1.8 && && & 0.059&23.6 && && && && && &&\\ \hline
\end{tabular}
\end{center}
\caption{Double differential invariant cross section $f(x_F,p_T)$ [mb/(GeV$^2$/c$^3$)]
         for $\pi^+$ in p+C interactions at 158 GeV/c. The statistical uncertainty 
	 $\Delta f$ is given in \%}
\label{tab:pip}
\end{table}
 
\begin{table}
\scriptsize
\begin{center}
\begin{tabular}{|c|c@{\hspace{2mm}}c@{\hspace{0.8mm}}
                  |c@{\hspace{2mm}}c@{\hspace{0.8mm}}
                  |c@{\hspace{2mm}}c@{\hspace{0.8mm}}
                  |c@{\hspace{2mm}}c@{\hspace{0.8mm}}
                  |c@{\hspace{2mm}}c@{\hspace{0.8mm}}
                  |c@{\hspace{2mm}}c@{\hspace{0.8mm}}
                  |c@{\hspace{2mm}}c@{\hspace{0.8mm}}
                  |c@{\hspace{2mm}}c@{\hspace{0.8mm}}
                  |c@{\hspace{2mm}}c@{\hspace{0.8mm}}|}
\hline
\multicolumn{19}{|c|}{$f(x_F,p_T), \Delta f$} \\
\hline
$p_T \backslash x_F$ & \multicolumn{2}{|c|}{-0.1} & \multicolumn{2}{|c|}{-0.075}  
                     & \multicolumn{2}{|c|}{-0.05} & \multicolumn{2}{|c|}{-0.04}  
                     & \multicolumn{2}{|c|}{-0.03} & \multicolumn{2}{|c|}{-0.025}
                     & \multicolumn{2}{|c|}{-0.02} & \multicolumn{2}{|c|}{-0.01}  
                     & \multicolumn{2}{|c|}{0.0} \\ \hline
0.05 &&  &&  &&  &&  &&  &&  &541.6&2.35 &646.3&2.19 &602.6&1.57\\ \hline
0.1 &&  &&  &&  &&  &525.3&2.11 &&  &595.9&1.74 &601.7&1.48 &579.8&1.43\\ \hline0.15 &&  &&  &&  &426.5&2.81 &481.8&1.88 &&  &511.6&1.69 &469.1&1.60 &498.6&1.39\\ \hline
0.2 &&  &&  &345.9&3.01 &364.0&2.74 &372.4&1.97 &&  &373.7&1.80 &374.9&1.68
    &367.2&1.66\\ \hline
0.25 &&  &&  &250.1&3.30 &275.0&2.53 &294.5&2.00 &&  &289.2&1.96 &285.9&1.91
     &296.1&1.78\\ \hline
0.3 &&  &160.4&1.77 &194.6&3.50 &197.4&2.89 &215.5&2.34 &&  &217.1&2.22 &210.9&2.18
    &214.0&2.06\\ \hline
0.4 & 86.2&3.33 &103.3&2.58 &112.8&2.75 &115.3&2.49 &115.2&2.09 &&  &110.7&2.17
    &118.6&2.03 &125.7&1.75\\ \hline
0.5 & 52.8&3.62 & 55.8&3.24 & 64.6&3.53 & 63.0&3.29 & 63.3&2.80 &&  &64.7&2.54
    & 63.5&3.03 & 64.4&2.50\\ \hline
0.6 & 31.2&3.99 & 31.7&4.10 & 32.7&4.91 & 35.5&3.61 & 36.6&3.43 &&  &34.2&3.47
    & 35.5&3.35 & 32.1&3.51\\ \hline
0.7 & 18.5&5.01 & 19.4&5.14 & 19.0&5.45 &19.31&4.79 &20.88&4.44 &&  &20.31&4.44
    &20.02&4.42 &20.30&4.48\\ \hline
0.8 &10.64&6.32 &10.38&6.41 &11.76&4.68 &&  &&  &11.74&3.77 &&  &&  &11.12&3.81\\ \hline
0.9 & 5.98&7.84 & 6.63&7.38 & 6.64&5.63 &&  &&  & 6.66&4.98 &&  &&  &6.72&4.95\\ \hline
1.0 & 3.84&8.72 & 3.83&8.14 & 3.79&7.56 &&  &&  & 4.21&6.32 &&  &&  &3.78&6.43\\ \hline
1.2 & 1.29&10.2 & 1.46&9.30 & 1.26&8.87 &&  &&  & 1.30&7.78 &&  &&  &1.246&8.62\\ \hline
1.4 & 0.491&10.6 &&  & 0.506&8.94 &&  &&  &&  &&  &&  &0.582&8.21\\ \hline
1.6 & 0.205&15.8 &&  & 0.189&14.8 &&  &&  &&  &&  &&  &0.184&14.3\\ \hline
1.8 & 0.093&21.4 &&  & 0.101&19.7 &&  &&  &&  &&  &&  &0.075&22.4\\ \hline
\hline
$p_T \backslash x_F$ & \multicolumn{2}{|c|}{0.01} & \multicolumn{2}{|c|}{0.02}
                     & \multicolumn{2}{|c|}{0.025} & \multicolumn{2}{|c|}{0.03}
                     & \multicolumn{2}{|c|}{0.04} & \multicolumn{2}{|c|}{0.05}
                     & \multicolumn{2}{|c|}{0.06} & \multicolumn{2}{|c|}{0.075}
                     & \multicolumn{2}{|c|}{0.1}\\ \hline
0.05 &576.1&1.71 &469.0&2.17 &&  &387.0&2.13 &336.6&2.15 &265.9&2.63 &225.5&3.11     
&200.6&2.60 &150.2&3.06\\ \hline
0.1 &539.5&1.58 &464.9&1.54 &&  &387.9&1.79 &316.2&1.93 &275.2&2.30 &246.7&2.58
    &205.1&2.28 &146.6&2.76\\ \hline
0.15 &445.0&1.50 &407.0&1.64 &&  &353.7&2.00 &314.0&1.64 &262.3&1.96 &238.0&2.17  
     &187.4&1.95 &148.8&2.18\\ \hline
0.2 &362.0&1.56 &320.5&1.69 &&  &294.7&1.98 &269.1&1.85 &230.8&1.87 &205.8&2.11
    &180.8&1.80 &131.7&2.14\\ \hline
0.25 &274.0&1.75 &256.7&1.86 &&  &235.6&2.03 &210.3&1.85 &192.3&2.15 &171.8&2.41  
     &147.1&1.78 &109.4&1.99\\ \hline
0.3 &218.1&2.01 &189.2&2.07 &&  &173.0&2.26 &163.8&2.53 &150.6&2.32 &132.1&2.64
    &125.0&1.75 & 96.1&2.03\\ \hline
0.4 &113.0&1.93 &107.0&1.85 &&  &101.8&1.97 &&  & 91.7&1.45 &&  & 74.2&1.50
    & 62.1&1.59\\ \hline
0.5 & 61.0&2.58 & 59.1&2.66 &&  & 58.6&2.76 &&  & 52.5&1.99 &&  & 44.9&2.20
    &38.85&1.81\\ \hline
0.6 & 35.2&3.37 & 33.4&3.53 &&  & 28.5&3.91 &&  &28.46&2.61 &&  &26.50&2.65
    &23.85&2.22\\ \hline
0.7 &18.92&4.53 &18.74&4.68 &&  &17.40&4.91 &&  &17.13&3.32 &&  &15.10&3.77
    &13.10&2.66\\ \hline
0.8 &&  &&  &10.58&3.96 &&  &&  & 9.07&4.35 &&  & 8.98&4.65 & 7.62&3.43\\ \hline
0.9 &&  &&  & 6.44&5.14 &&  &&  & 5.64&6.23 &&  & 5.06&6.06 & 4.36&5.00\\ \hline
1.0 &&  &&  & 3.28&6.71 &&  &&  & 3.84&7.43 &&  & 2.83&9.23 & 2.63&6.22\\ \hline
1.2 &&  &&  & 1.376&7.42 &&  &&  & 1.03&9.86 &&  & 1.06&10.1 & 1.007&7.38\\ \hline
1.4 &&  &&  &&  &&  &&  & 0.440&10.6 &&  &&  & 0.449&11.0\\ \hline
1.6 &&  &&  &&  &&  &&  & 0.150&18.3 &&  &&  & 0.197&16.3\\ \hline
1.8 &&  &&  &&  &&  &&  & 0.050&36.2 &&  &&  & 0.064&28.7\\ \hline
\hline
$p_T \backslash x_F$ & \multicolumn{2}{|c|}{0.125} & \multicolumn{2}{|c|}{0.15}
                     & \multicolumn{2}{|c|}{0.2} & \multicolumn{2}{|c|}{0.25}
                     & \multicolumn{2}{|c|}{0.3} & \multicolumn{2}{|c|}{0.4}
                     & \multicolumn{2}{|c|}{0.5} & \multicolumn{2}{|c|}{}
                     & \multicolumn{2}{|c|}{}\\ \hline
0.05 &111.9&3.92 & 86.1&3.94 & 55.1&5.14 & 41.4&6.81 & 33.3&8.22 & 12.1&13.6 &&
 &&
     && \\ \hline
0.1 &117.8&3.44 & 81.8&3.66 & 58.1&4.55 & 41.8&4.75 & 29.4&7.69 & 10.8&10.2
    &5.1&16.2 && &&\\ \hline
0.15 &111.0&2.92 & 83.1&2.86 & 51.4&3.80 &&  &&  &&  &&  &&  && \\ \hline
0.2 & 98.8&2.64 & 81.9&2.54 & 53.5&3.27 & 30.7&3.96 & 22.7&6.11 &  9.6&7.61
    & 4.34&12.4 && &&\\ \hline
0.25 & 90.7&2.42 & 72.7&2.41 & 42.6&3.28 &&  &&  &&  &&  &&  && \\ \hline
0.3 & 73.4&2.57 & 58.2&2.49 & 39.3&3.09 & 25.9&3.72 & 17.2&5.71 & 7.69&6.83
    & 3.23&11.8 && &&\\ \hline
0.4 & 50.5&1.92 &42.19&1.79 &28.69&2.55 &18.31&3.77 &12.21&5.83 & 5.33&7.03
    & 2.10&13.0 && &&\\ \hline
0.5 &31.78&2.23 &27.36&2.03 &21.15&2.72 &14.13&3.93 & 9.79&5.88 & 3.83&7.61
    & 1.57&13.3 && &&\\ \hline
0.6 &19.02&2.67 &17.09&2.71 &12.00&3.30 & 9.28&4.51 & 6.12&6.83 & 3.07&7.43
    &0.97&15.1 && &&\\ \hline
0.7 &&  &10.30&2.99 &7.40&4.17 &&  &&  &&  &&  &&  &&\\ \hline
0.8 &&  &5.97&3.68 &5.12&4.74 &3.75&4.29 &2.60&6.45 &1.20&7.56 &0.56&12.0
    && &&\\ \hline
0.9 &&  &3.60&4.57 &2.65&6.31 &&  &&  &&  &&  &&  &&\\ \hline
1.0 &&  &2.28&5.58 &1.73&7.50 &1.45&6.22 &0.95&9.62 &0.588&9.57 &0.189&15.2
    && &&\\ \hline
1.2 &&  &0.727&7.42 &0.674&7.96 &0.540&9.49 &0.392&13.7 &0.173&16.9
    &0.077&27.1 && &&\\ \hline
1.4 &&  &&  &0.293&9.42 &&  &0.150&14.3 &0.063&25.7 &0.031&39.6 &&
    &&\\ \hline
1.6 &&  &&  &0.106&14.0 &&  &0.064&21.6 &0.026&32.3 &&  &&  &&\\
\hline
1.8 &&  &&  &0.037&24.4 &&  &&  &&  &&  &&  &&\\ \hline
\end{tabular}
\end{center}
\caption{Double differential invariant cross section $f(x_F,p_T)$ [mb/(GeV$^2$/c$^3$)]
         for $\pi^-$ in p+C interactions at 158 GeV/c. The statistical uncertainty 
         $\Delta f$ is given in \%}
\label{tab:pim}
\end{table}

%
%
\subsection{$\pi^+/\pi^-$ ratios}
\begin{figure}[t!]
  \centering
  \epsfig{file=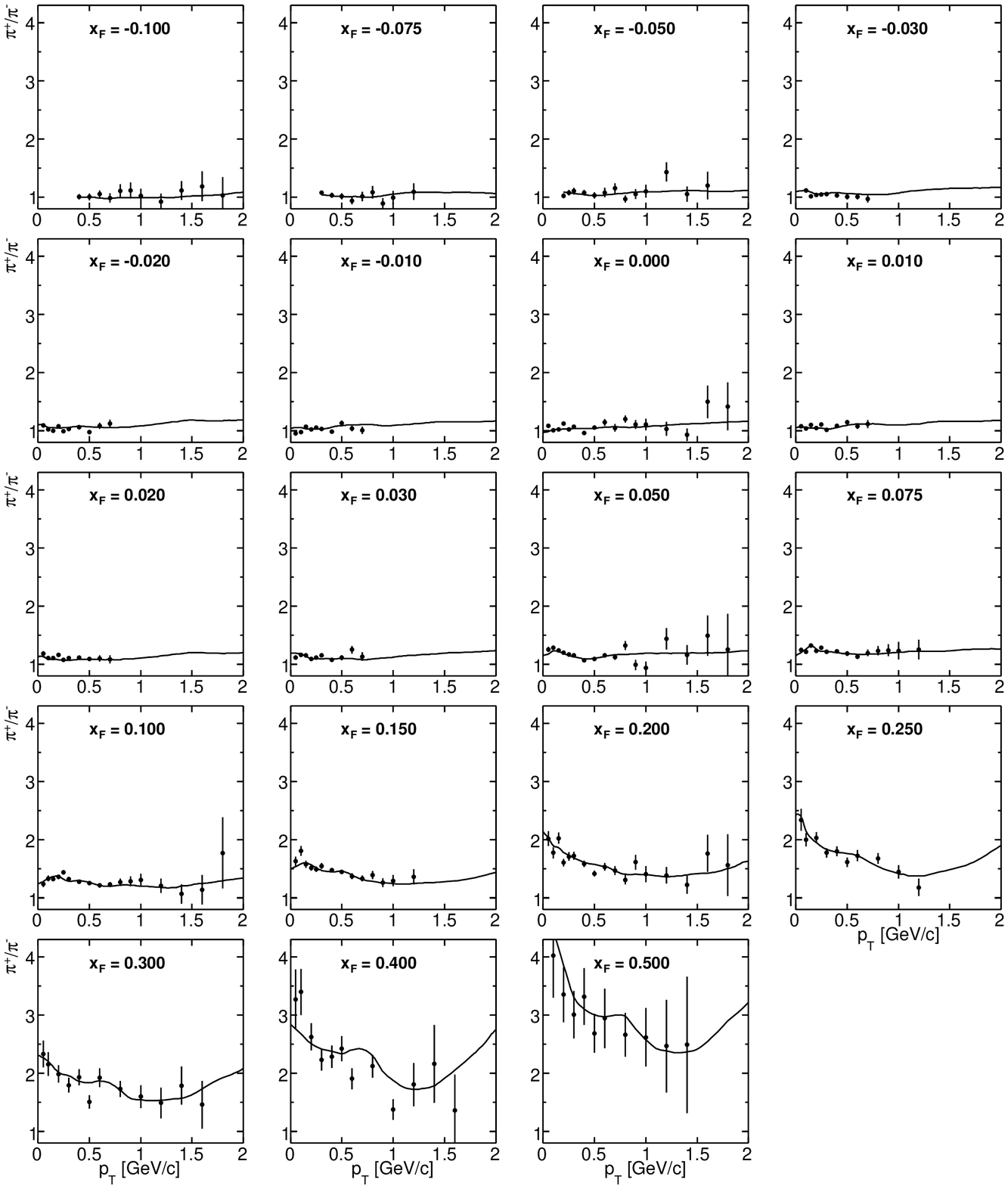,width=16cm}
  \caption{Ratio of invariant cross section for $\pi^+$ and $\pi^-$ as a function
           of $p_T$ at fixed $x_F$. The lines represent the result of the data 
	   interpolation}
  \label{fig:pt_ratio}
\end{figure}

\begin{figure}[t]
  \centering
  \epsfig{file=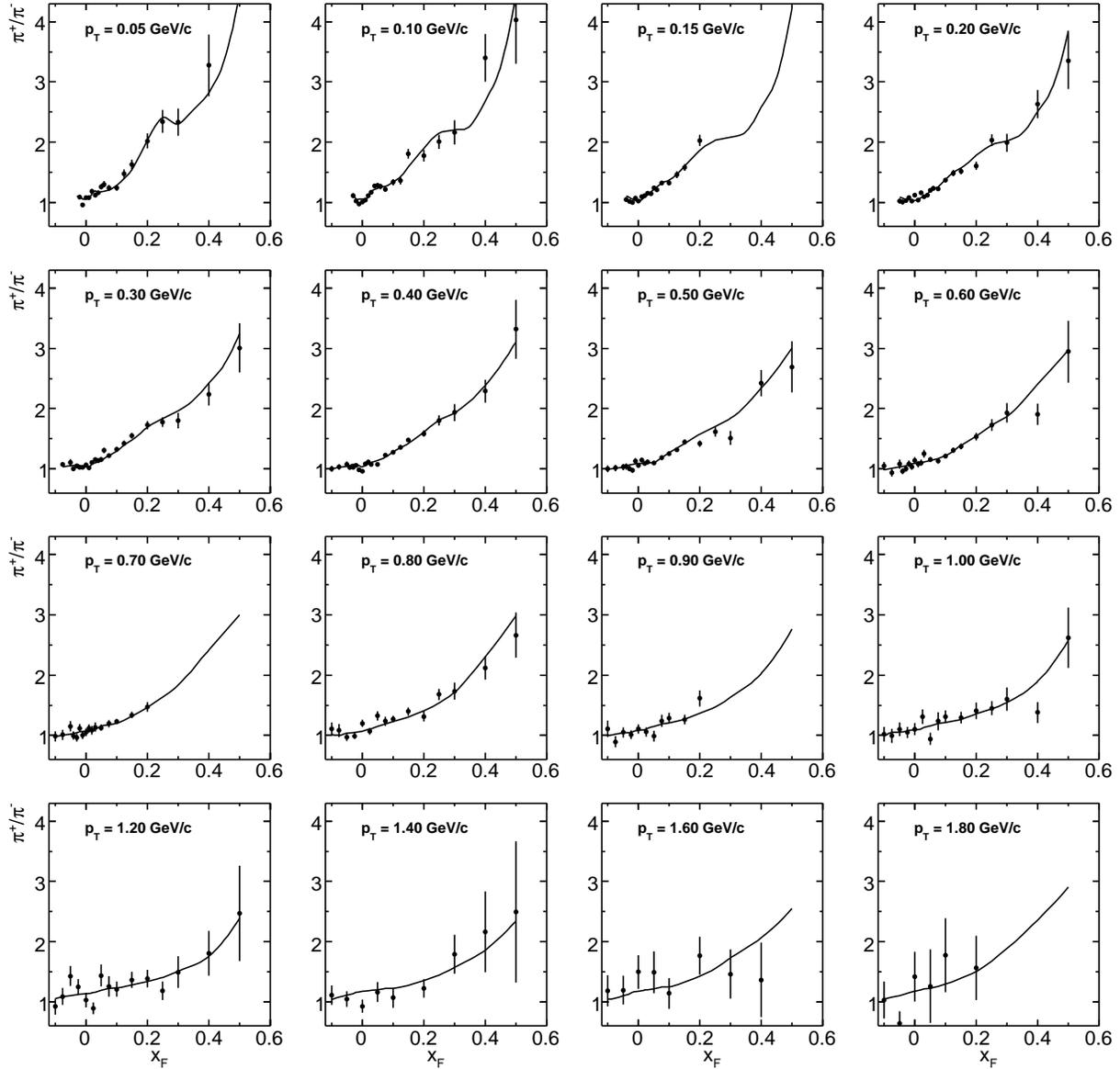,width=16cm}
  \caption{Ratio of invariant cross section for $\pi^+$ and $\pi^-$ as a function
           of $x_F$ at fixed $p_T$}
  \label{fig:xf_ratio}
\end{figure}

\vspace{3mm}
The dependence of the $\pi^+/\pi^-$ ratio on $p_T$ and $x_F$ carries important
information concerning the details of charge conservation in the
hadronization process. As presented in Figs.~\ref{fig:pt_ratio} and 
\ref{fig:xf_ratio}, important substructures in both variables become 
visible which are directly comparable to the situation in p+p interactions 
\cite{bib:pp_paper}.

\clearpage

This comparison shows global similarities. Remarkable differences
are however visible in the backward direction where the ratios are expected
to approach unity due to the isoscalar nature of the target nucleus,
and in the detailed structures in the projectile hemisphere.

%
%
\subsection{Rapidity and transverse mass distributions}
\begin{figure}[h]
  \centering
  \epsfig{file=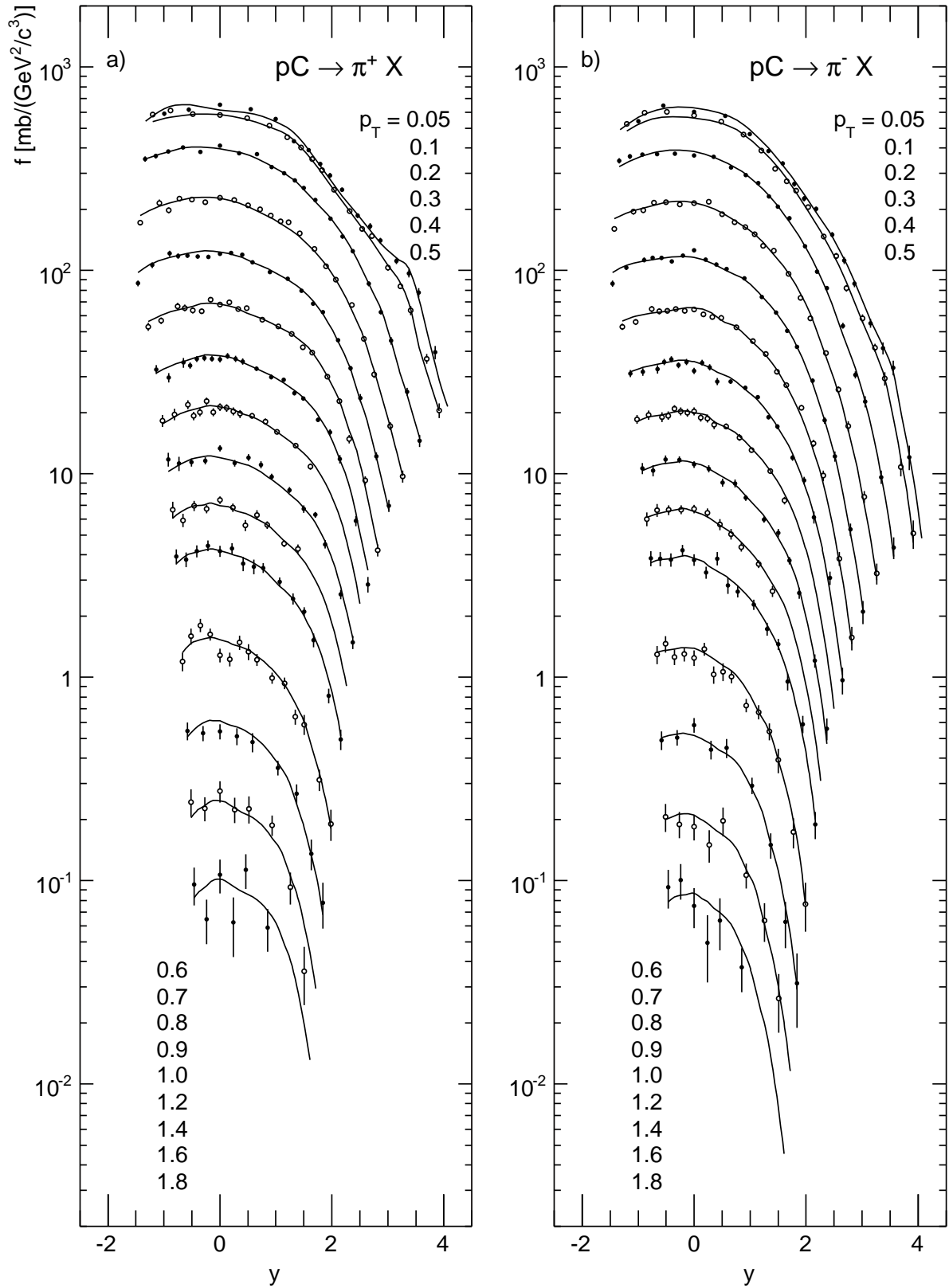,width=16cm}
  \caption{Invariant cross section as a function of $y$ at fixed $p_T$ for a) $\pi^+$
           and b) $\pi^-$ produced in p+C collisions at 158~GeV/c}
  \label{fig:rap_dist}
\end{figure}

\vspace{3mm}
The rapidity distributions at fixed $p_T$ presented in
Fig.~\ref{fig:rap_dist} extend up to 1.5 units into the target hemisphere 
at $p_T<$~0.4~GeV/c and therefore allow a clear view of the asymmetry 
which reaches about 0.25 units at low $p_T$. 

\begin{figure}[h]
  \centering
  \epsfig{file=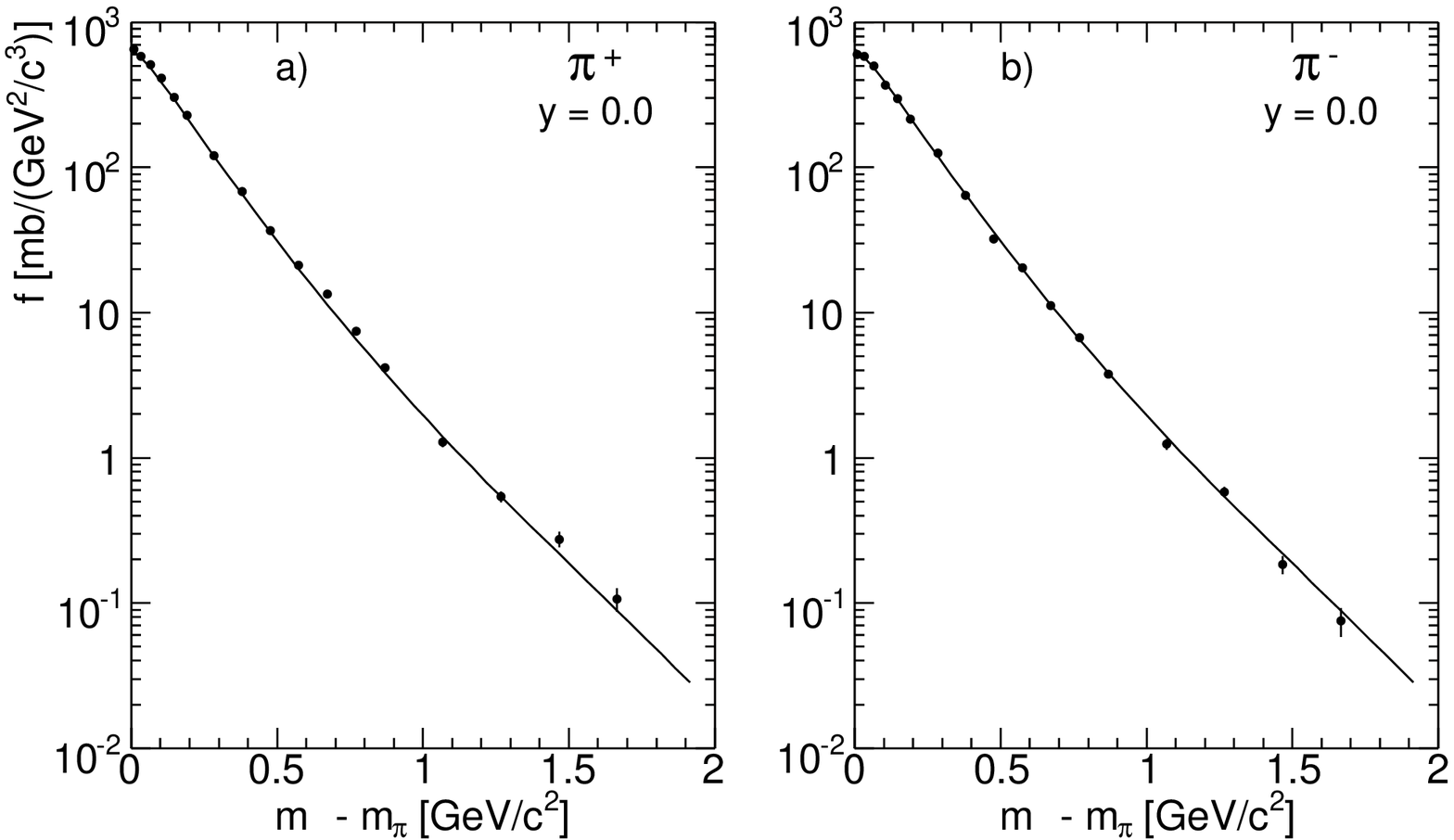,width=15cm}
  \caption{Invariant cross section as a function of $m_T - m_{\pi}$ for a) $\pi^+$ and
           b) $\pi^-$ produced at $y$~= 0.0 }
  \label{fig:slope}
\end{figure}

In comparison to p+p collisions \cite{bib:pp_paper}, they show an 
important steepening in the projectile hemisphere at all transverse momenta, 
whereas the characteristic deformation at low $p_T$ and $y$ is still 
clearly visible. The $m_T$ distributions for both charges are shown 
in Fig.~\ref{fig:slope}.

\begin{figure}[h]
  \centering
  \epsfig{file=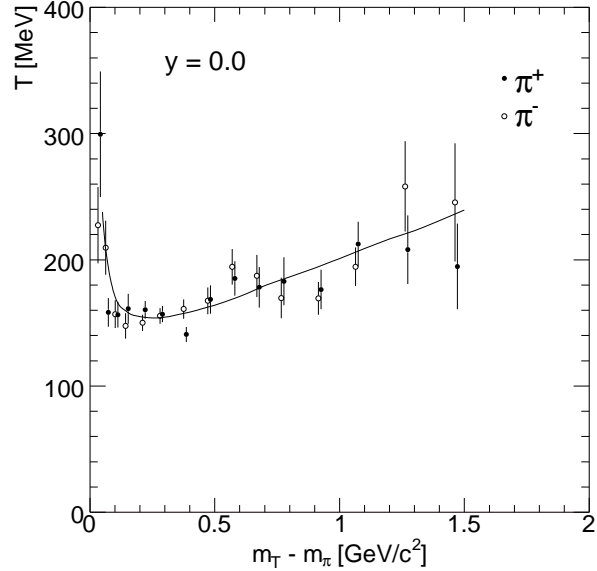,width=8cm}
  \caption{Local slope of the $m_T$ distribution as a function of $m_T - m_{\pi}$
           for $\pi^+$ and $\pi^-$. The line shown is to guide the eye }
  \label{fig:mt-slope}
\end{figure}
 
As stressed in \cite{bib:pp_paper} these distributions are all
but exponential and the variation of the local inverse slopes $T$ with 
$m_T$, shown in Fig.~\ref{fig:mt-slope}, is even more pronounced than in p+p 
collisions.

%
%
\section{Comparison to other data}
\begin{figure}[t]
  \centering
  \epsfig{file=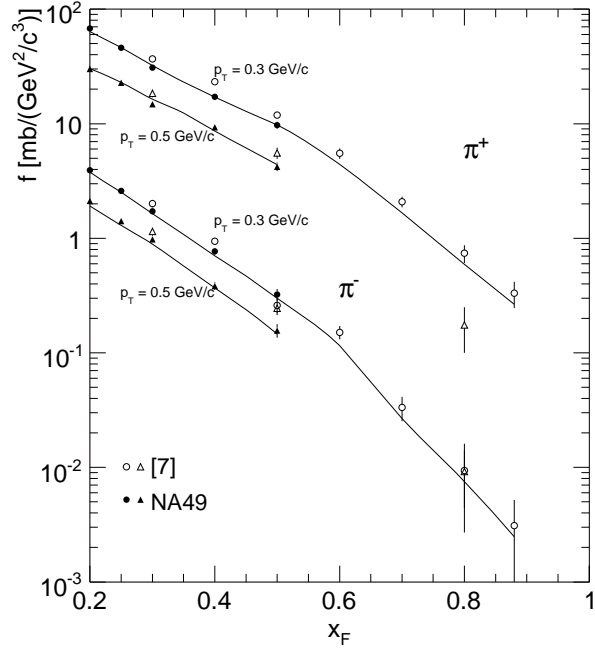,width=8cm}
  \caption{Comparison of the invariant cross section as a function of $x_F$ at 
           fixed $p_T$ from NA49 (full symbols) with measurement from 
           \cite{bib:barton} (open symbols). The $\pi^-$ data lines are 
	   multiplied by 0.1 to allow a separation from the $\pi^+$ }
  \label{fig:barton}
\end{figure}

\vspace{3mm}
As shown in Sect.~\ref{sec:exp_sit}, the only existing data set which can be
directly compared to the NA49 results is that of Barton et al.\cite{bib:barton}. 
This comparison is shown in Fig.~\ref{fig:barton}.

For the 10 points which overlap with the NA49 data a clear upward
deviation with an average of +25\% or 3.6 standard deviations is
evident. This deviation is somewhat hard to understand as these
results come from a group which has published results on p+p
interactions with the same apparatus \cite{bib:brenner} which show excellent 
agreement with NA49 (see \cite{bib:pp_paper} for a detailed discussion). Also
the p+p reference data obtained in the framework of \cite{bib:barton} are 
internally consistent, notwithstanding their sizable statistical
errors, both with \cite{bib:brenner} and with NA49. Preliminary analysis
reveals the same problem also with proton yields in p+C collisions.

\begin{figure}[h]
  \centering
  \epsfig{file=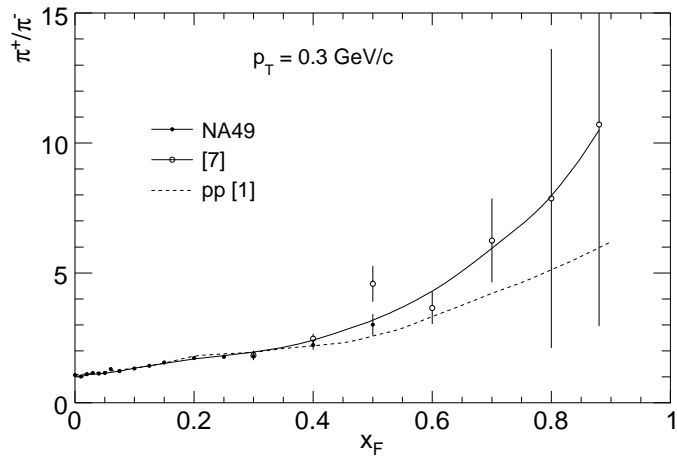,width=9cm}
  \caption{Comparison of the $\pi^+/\pi^-$ ratio as a function of $x_F$ at fixed
           $p_T$ from NA49 (full circles) with measurement from 
           \cite{bib:barton} (open circles). The dashed line represents
           the p+p data \cite{bib:pp_paper} }
  \label{fig:barton_rat}
\end{figure}

As the differences are similar for both pion charges, the $\pi^+/\pi^-$
ratios are expected to be unaffected by the problem. This is indeed
the case, the good agreement between the two data sets at $p_T$~= 0.3~GeV/c
as a function of $x_F$ is shown in Fig.~\ref{fig:barton_rat}.

Since the Barton \cite{bib:barton} data extend up to $x_F$ = 0.88 at their 
measured transverse momenta, one can use the consistency of the $\pi^+/\pi^-$ 
ratio to extend the NA49 data - albeit with large error bars - into the
region of large $x_F$. This shows, as indicated in Fig.~\ref{fig:barton_rat}, 
a sizable increase of the charge ratio with respect to the p+p data also
given in the Fig. It is also possible, under the 
assumption that the large systematic yield difference has no dependence on
$p_T$ or $x_F$, to extend the interpolation of the NA49 data towards
large $x_F$ by imposing a 25\% reduction on the Barton et al. data.
This extension is also indicated in Fig.~\ref{fig:barton} for   
$p_T$~= 0.3~GeV/c. It shows the characteristic break in the $x_F$
dependence at $x_F$ between 0.5 and 0.6 also visible in the p+p data 
\cite{bib:pp_paper}.

%
%
\section{Integrated distributions}
\begin{figure}[h]
  \centering
  \epsfig{file=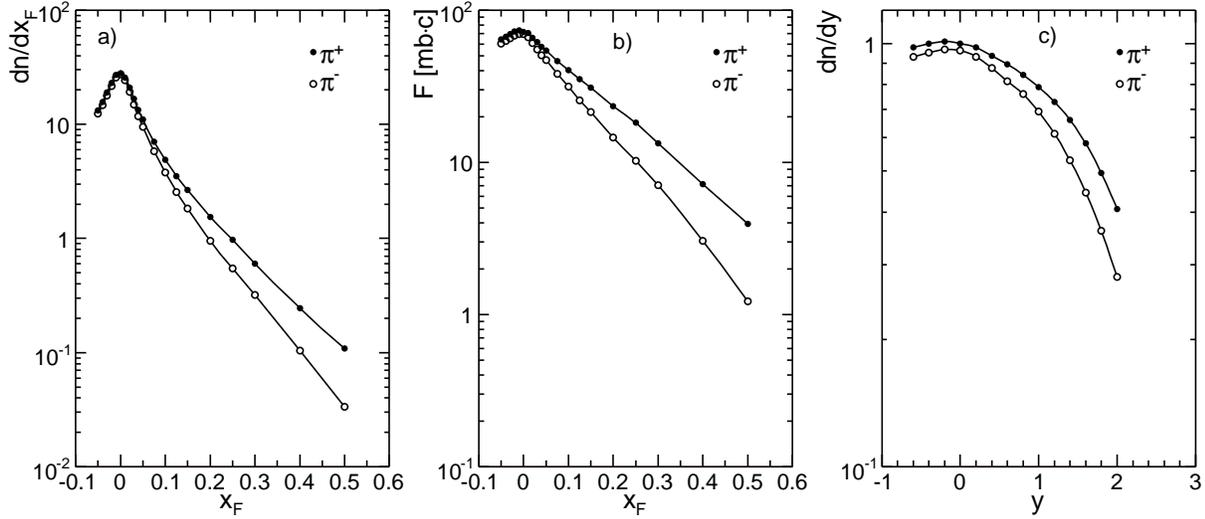,width=16cm}
  \caption{Integrated distributions of $\pi^+$ and $\pi^-$ produced in p+C 
           interactions at 158~GeV/c:
           a) density distribution $dn/dx_F$ as a function of $x_F$;
           b) invariant cross section $F$ as a function of $x_F$;
           c) density distribution $dn/dy$ as a function of $y$}
  \label{fig:integr}
\end{figure}

\vspace{3mm}
The $p_T$ integrated yields


\begin{align}
  dn/dx_F &= \pi/\sigma_{\textrm{inel}} \cdot \sqrt{s}/2 \cdot 
             \int{f/E \cdot dp_T^2} \nonumber \\
  F &= \int{f \cdot dp_T^2}  \\
  dn/dy &= \pi/\sigma_{\textrm{inel}} \cdot \int{f \cdot dp_T^2} \nonumber  
\end{align}
are obtained from the interpolated data. The numerical values are
given in Table~\ref{tab:integr} and presented as functions of $x_F$ and 
$y$ in Fig.~\ref{fig:integr}.

\vspace{5mm}
\begin{table}[h]
\scriptsize
\renewcommand{\tabcolsep}{0.10pc} 
\renewcommand{\arraystretch}{1.05} 
\begin{center}
\begin{tabular}{|c|cc|cc|cc|cc||cc|cc|cc|cc||c|c|c|}
\hline
 & \multicolumn{8}{|c||}{$\pi^+$} &\multicolumn{8}{|c||}{$\pi^-$} & & $\pi^+$ & $\pi^-$ \\
\hline
 $x_F$& $F$& $\Delta$& $dn/dx_F$& $\Delta$& $\langle p_T \rangle$& $\Delta$
      & $\langle p_T^2 \rangle$ & $\Delta$
      & $F$& $\Delta$& $dn/dx_F$& $\Delta$& $\langle p_T \rangle$& $\Delta$
      & $\langle p_T^2 \rangle $& $\Delta$
      & $y$& $dn/dy$ & $dn/dy$ \\ \hline
 -0.05  & 64.312 & 0.57 & 13.211  & 0.57 & 0.3267 & 0.40 & 0.1533 & 0.80 &
          60.470 & 0.57 & 12.401  & 0.57 & 0.3287 & 0.40 & 0.1543 & 0.80 &
 -0.6   & 0.9795 & 0.9311 \\         
 -0.04  & 66.736 & 0.52 & 15.726  & 0.52 & 0.3114 & 0.31 & 0.1411 & 0.60 &
          62.771 & 0.52 & 14.749  & 0.52 & 0.3141 & 0.31 & 0.1425 & 0.60 &
 -0.4   & 0.9994 & 0.9521 \\
 -0.03  & 69.546 & 0.48 & 19.030  & 0.48 & 0.2954 & 0.30 & 0.1287 & 0.55 &
          65.076 & 0.48 & 17.768  & 0.48 & 0.2970 & 0.31 & 0.1294 & 0.55 &
 -0.2   & 1.0120 & 0.9688 \\ 
 -0.02  & 72.264 & 0.38 & 23.052  & 0.38 & 0.2801 & 0.28 & 0.1177 & 0.52 &
          68.061 & 0.38 & 21.743  & 0.38 & 0.2796 & 0.30 & 0.1168 & 0.55 &
  0.0   & 1.0010 & 0.9646 \\
 -0.01  & 73.693 & 0.34 & 26.993  & 0.34 & 0.2658 & 0.27 & 0.1077 & 0.50 &
          69.751 & 0.33 & 25.747  & 0.33 & 0.2631 & 0.30 & 0.1049 & 0.54 &
  0.2   & 0.9796 & 0.9317 \\
  0.0   & 71.923 & 0.32 & 28.088  & 0.32 & 0.2586 & 0.27 & 0.1030 & 0.48 &
          69.484 & 0.30 & 27.364  & 0.30 & 0.2558 & 0.30 & 0.1008 & 0.54 &
  0.4   & 0.9373 & 0.8751 \\
  0.01  & 70.586 & 0.32 & 25.756  & 0.32 & 0.2672 & 0.27 & 0.1088 & 0.50 &
          66.470 & 0.31 & 24.424  & 0.31 & 0.2648 & 0.30 & 0.1062 & 0.55 &
  0.6   & 0.8941 & 0.8159 \\
  0.02  & 66.086 & 0.32 & 20.903  & 0.32 & 0.2840 & 0.27 & 0.1210 & 0.50 &
          60.877 & 0.32 & 19.247  & 0.32 & 0.2844 & 0.30 & 0.1206 & 0.55 &
  0.8   & 0.8448 & 0.7597 \\
  0.03  & 61.594 & 0.33 & 16.683  & 0.33 & 0.3012 & 0.29 & 0.1339 & 0.55 &
          55.378 & 0.35 & 14.938  & 0.35 & 0.3042 & 0.31 & 0.1352 & 0.58 &
  1.0   & 0.7898 & 0.6905 \\
  0.04  & 57.541 & 0.35 & 13.389  & 0.35 & 0.3205 & 0.30 & 0.1491 & 0.60 &
          50.685 & 0.37 & 11.757  & 0.37 & 0.3237 & 0.32 & 0.1506 & 0.62 &
  1.2   & 0.7291 & 0.6131 \\
  0.05  & 54.137 & 0.35 & 10.981  & 0.35 & 0.3379 & 0.32 & 0.1637 & 0.62 &
          47.027 & 0.38 &  9.516  & 0.38 & 0.3413 & 0.32 & 0.1649 & 0.62 &
  1.4   & 0.6611 & 0.5296 \\
  0.075 & 46.426 & 0.35 &  7.067  & 0.35 & 0.3734 & 0.32 & 0.1957 & 0.62 &
          38.142 & 0.38 &  5.790  & 0.38 & 0.3781 & 0.32 & 0.1993 & 0.62 &
  1.6   & 0.5807 & 0.4451 \\
  0.1   & 40.396 & 0.35 &  4.904  & 0.35 & 0.3960 & 0.30 & 0.2182 & 0.58 &
          31.456 & 0.38 &  3.803  & 0.38 & 0.4045 & 0.32 & 0.2268 & 0.62 &
  1.8   & 0.4942 & 0.3612 \\      
  0.125 & 35.101 & 0.37 &  3.531  & 0.37 & 0.4144 & 0.32 & 0.2375 & 0.65 &
          25.590 & 0.42 &  2.561  & 0.42 & 0.4277 & 0.33 & 0.2525 & 0.65 &
  2.0   & 0.4069 & 0.2806 \\      
  0.15  & 30.875 & 0.38 &  2.648  & 0.38 & 0.4267 & 0.32 & 0.2520 & 0.65 &
          21.385 & 0.47 &  1.826  & 0.47 & 0.4436 & 0.34 & 0.2708 & 0.70 &
        & & \\
  0.2   & 23.488 & 0.50 &  1.5499 & 0.50 & 0.4489 & 0.37 & 0.2797 & 0.70 &
          14.535 & 0.62 &  0.9555 & 0.62 & 0.4730 & 0.37 & 0.3053 & 0.70 &
        & & \\
  0.25  & 18.174 & 0.65 &  0.9734 & 0.65 & 0.4607 & 0.41 & 0.2967 & 0.80 &
          10.258 & 0.87 &  0.5475 & 0.87 & 0.4909 & 0.43 & 0.3308 & 0.85 &
        & & \\
  0.3   & 13.327 & 0.87 &  0.5996 & 0.87 & 0.4755 & 0.44 & 0.3154 & 0.90 &
           7.105 & 1.15 &  0.3190 & 1.15 & 0.4993 & 0.53 & 0.3425 & 1.10 &
        & & \\
  0.4   &  7.207 & 1.15 &  0.2455 & 1.15 & 0.4912 & 0.61 & 0.3344 & 1.10 &
           3.060 & 1.60 &  0.1041 & 1.60 & 0.5159 & 0.63 & 0.3664 & 1.15 &
        & & \\
  0.5   &  3.943 & 1.60 &  0.1081 & 1.60 & 0.4714 & 0.90 & 0.3157 & 1.60 &
           1.226 & 2.40 &  0.0336 & 2.40 & 0.5125 & 1.00 & 0.3620 & 2.00 &  
        & & \\ \hline
\end{tabular}
\end{center}
\vspace{-2mm}
\caption{$p_T$ integrated invariant cross section $F$ [mb$\cdot$c],
         density distribution $dn/dx_F$, mean transverse momentum $\langle p_T \rangle $
         [GeV/c], mean transverse momentum squared $\langle p_T^2 \rangle $ 
         [(GeV/c)$^2$] as a function of $x_F$, as well as density distribution 
         $dn/dy$ as a function of $y$ for $\pi^+$ and $\pi^-$. The statistical 
         uncertainty $\Delta$ for each quantity is given in \% }
\label{tab:integr}
\end{table}
\vspace{5mm}

\begin{figure}[h]
  \centering
  \epsfig{file=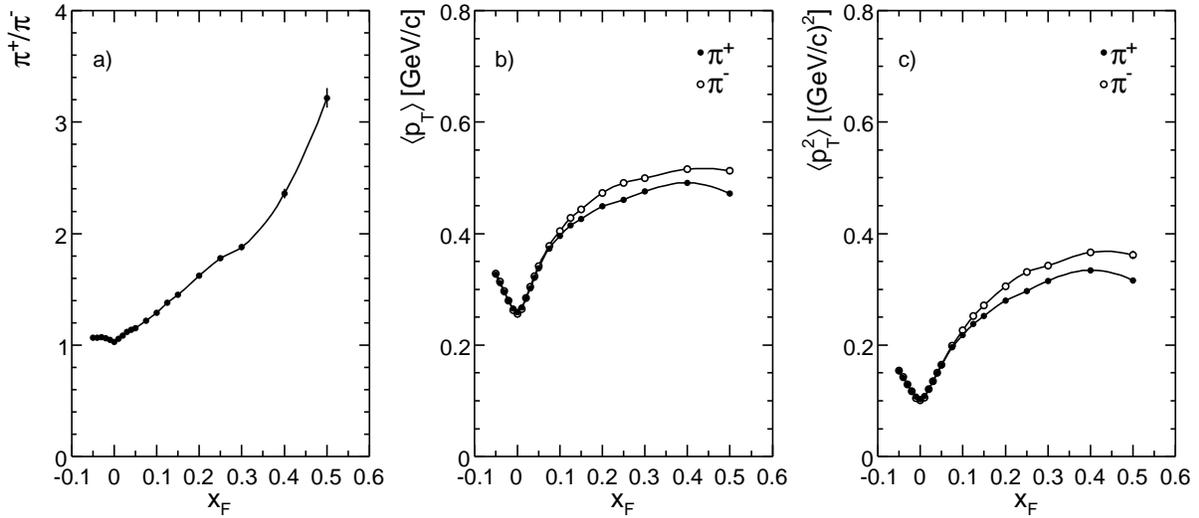,width=16cm}
  \caption{a) $\pi^+/\pi^-$ ratio, b) mean $p_T$, and c) mean $p_T^2$ as a function 
           of $x_F$ for $\pi^+$ and $\pi^-$ produced in p+C interactions at 158~GeV/c}
  \label{fig:mean}
\end{figure}

The $p_T$ integrated $\pi^+/\pi^-$ ratios, and the first and second moments of the
$p_T$ distributions are shown as functions of $x_F$ in Fig.~\ref{fig:mean}.

Due to the absence of data with sufficient phase space coverage 
allowing for integration over transverse momentum, a comparison with
other experiments is unfortunately not possible.

%
%
\section{Dependence on the number of grey protons}
\vspace{3mm}
\label{sec:ncd}
Due to the large fraction of single projectile collisions in minimum
bias p+C interactions, the number of grey protons 
$n_{\textrm{grey}}$ measured in this
reaction has a steep maximum at zero counts, as shown in Fig.~\ref{fig:ncd} above.
Correspondingly the event sample decreases rapidly from 377~000 to
102~000 and 26~000 respectively by selecting events with one or two grey 
protons. A general study of double differential inclusive cross sections
as described above is therefore not feasible for these
subsamples with reasonable statistical errors. On the other hand 
grey proton selection allows an important extension of the physics
analysis since it effectively suppresses the contribution of peripheral
collisions in favour of more central events with multiple projectile
collisions.

\begin{figure}[t]
  \centering
  \epsfig{file=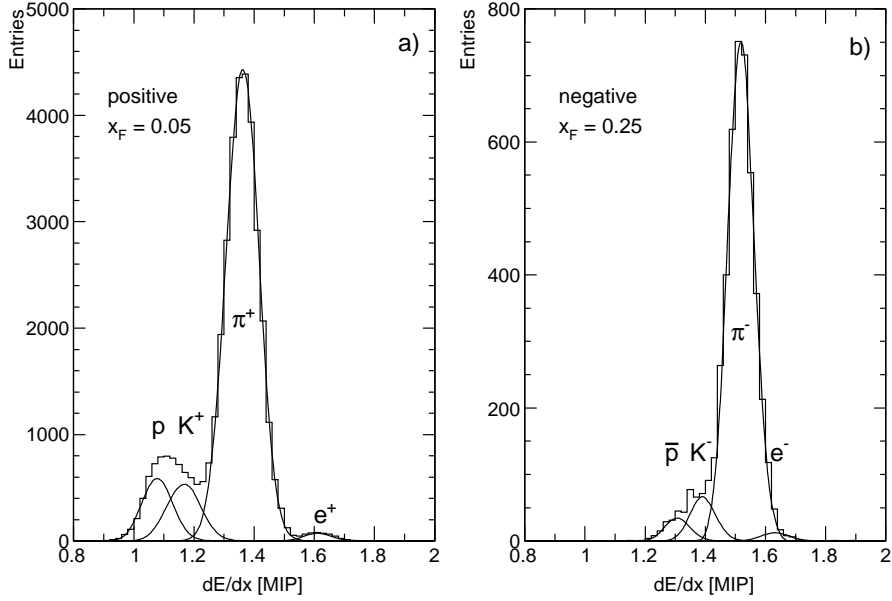,width=12cm}
  \caption{$dE/dx$ for $p_T$ integrated fits for a) $\pi^+$ and b) $\pi^-$}
  \label{fig:interg_fit}
\end{figure}

The problem of limited statistics can be overcome by extracting the $p_T$ 
integrated yields directly from the data sample as a function of $x_F$ only.
There is a limit to this procedure set by the necessities of
particle identification. The variation of the total momentum 
over the chosen bin width in $x_F$ must be small enough to allow for
the necessary resolution of the energy loss distribution. At low $x_F$
this variation becomes prohibitive due to the strong variation of the
total momentum with $p_T$. The method is therefore limited to 
$x_F\geq$~0.025. The quality of the $dE/dx$ distributions obtained 
is shown in Fig.~\ref{fig:interg_fit} for two values of $x_F$.

\begin{table}[b]
\footnotesize
\renewcommand{\tabcolsep}{0.14pc} 
\renewcommand{\arraystretch}{1.05} 
\begin{center}
\begin{tabular}{|c|cc|cc|cc||cc|cc|cc|}
\hline
       & \multicolumn{6}{|c||}{$\pi^+$} & \multicolumn{6}{|c|}{$\pi^-$} \\
\hline
       & \multicolumn{2}{|c|}{minimum bias} 
       & \multicolumn{2}{|c|}{$n_{\textrm{grey}}\geq$~1}
       & \multicolumn{2}{|c||}{$n_{\textrm{grey}}\geq$~2} 
       & \multicolumn{2}{|c|}{minimum bias} 
       & \multicolumn{2}{|c|}{$n_{\textrm{grey}}\geq$~1}
       & \multicolumn{2}{|c|}{$n_{\textrm{grey}}\geq$~2} \\
\hline
 $x_F$ & $dn/dx_F$ & $\Delta$ & $dn/dx_F$ & $\Delta$ & $dn/dx_F$ & $\Delta$
       & $dn/dx_F$ & $\Delta$ & $dn/dx_F$ & $\Delta$ & $dn/dx_F$ & $\Delta$ \\
\hline
 0.025 & 18.920  & 0.7  & 21.198  & 1.2  & 22.872  &  2.3  
       & 17.175  & 0.7  & 18.968  & 1.3  & 21.157  &  2.5    \\       
 0.05  & 10.996  & 0.6  & 11.875  & 1.0  & 12.704  &  2.0
       &  9.469  & 0.5  & 10.509  & 0.9  & 11.179  &  1.6    \\ 
 0.10  &  4.908  & 0.6  &  5.045  & 1.1  &  5.270  &  2.2
       &  3.793  & 0.8  &  3.935  & 1.4  &  4.036  &  2.8    \\
 0.15  &  2.641  & 0.7  &  2.649  & 1.3  &  2.806  &  2.6
       &  1.810  & 1.0  &  1.848  & 1.9  &  1.961  &  3.6    \\
 0.20  &  1.536  & 1.0  &  1.463  & 1.8  &  1.510  &  3.6
       &  0.977  & 1.2  &  0.966  & 2.2  &  0.960  &  4.4    \\
 0.25  &  0.969  & 1.2  &  0.877  & 2.4  &  0.880  &  4.7
       &  0.540  & 1.6  &  0.556  & 3.0  &  0.547  &  6.0    \\
 0.30  &  0.592  & 1.5  &  0.549  & 3.0  &  0.526  &  6.1
       &  0.322  & 2.4  &  0.308  & 4.0  &  0.292  &  8.2    \\
 0.35  &  0.399  & 1.8  &  0.337  & 3.8  &  0.3032 &  8.1
       &  0.174  & 3.0  &  0.153  & 5.7  &  0.1872 & 10.2    \\
 0.40  &  0.243  & 2.3  &  0.216  & 4.8  &  0.1793 & 10.5
       &  0.107  & 3.5  &  0.0989 & 7.1  &  0.0933 & 14.5    \\
 0.50  &  0.0973 & 3.6  &  0.0785 & 8.0  &  0.0672 & 17.2 
       &  0.0337 & 6.2  &  0.0298 &12.9  &         &         \\
\hline 
\end{tabular}
\end{center}
\vspace{-2mm}
\caption{$dn/dx_F$ for $n_{\textrm{grey}}$ selection. The uncertainty 
         $\Delta$ is given in \% }
\label{tab:ngray}
\end{table}

\begin{figure}[t]
  \centering
  \epsfig{file=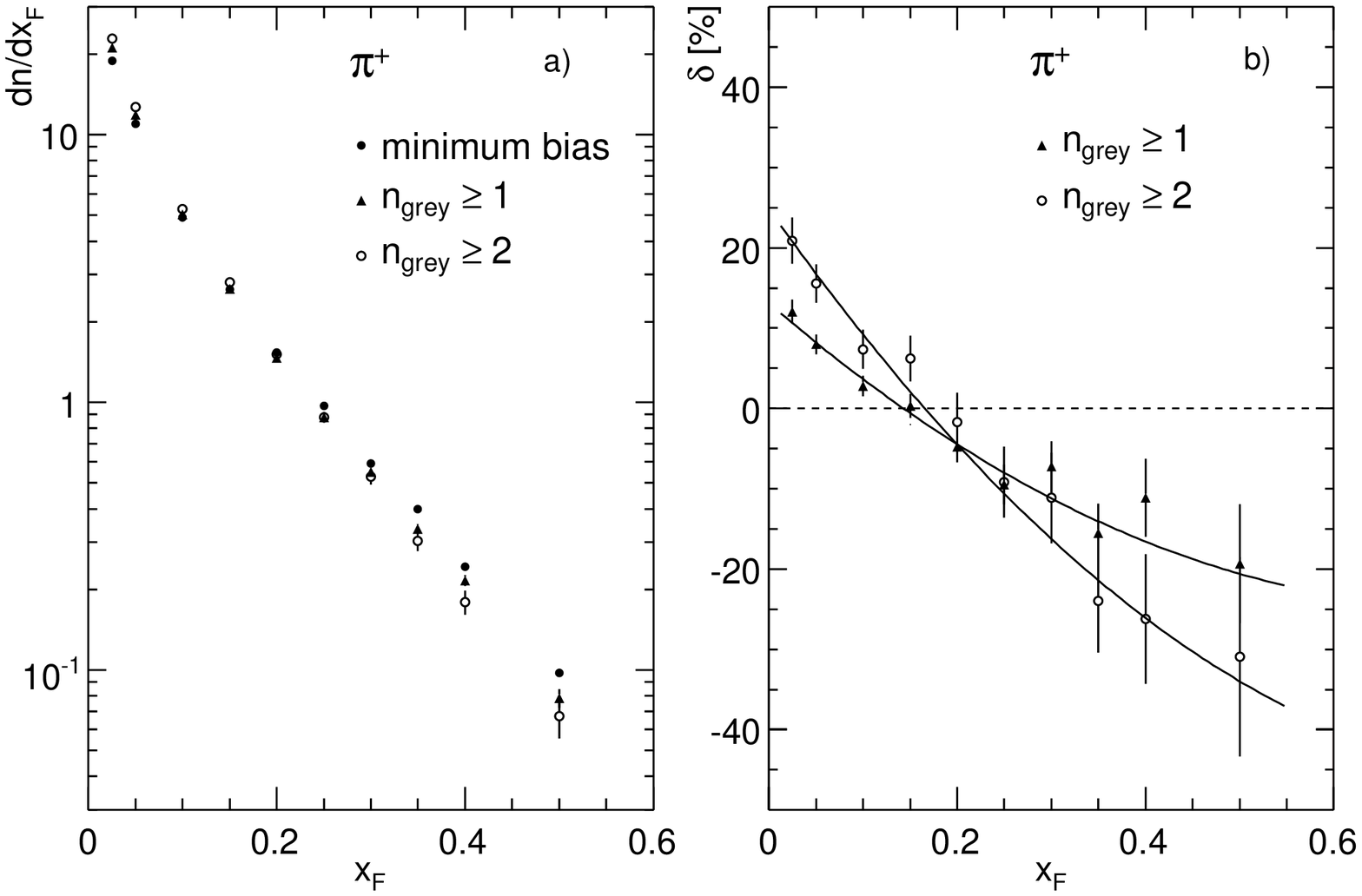,width=14cm}
  \caption{a) $dn/dx_F$ for $n_{\textrm{grey}}$ selection and 
           b) the difference $\delta$ between samples with 
	   $n_{\textrm{grey}}$ selection and minimum bias for $\pi^+$. 
	   The lines shown are to guide the eye }
  \label{fig:ngrey_pos}
\end{figure}

\begin{figure}[h]
  \centering
  \epsfig{file=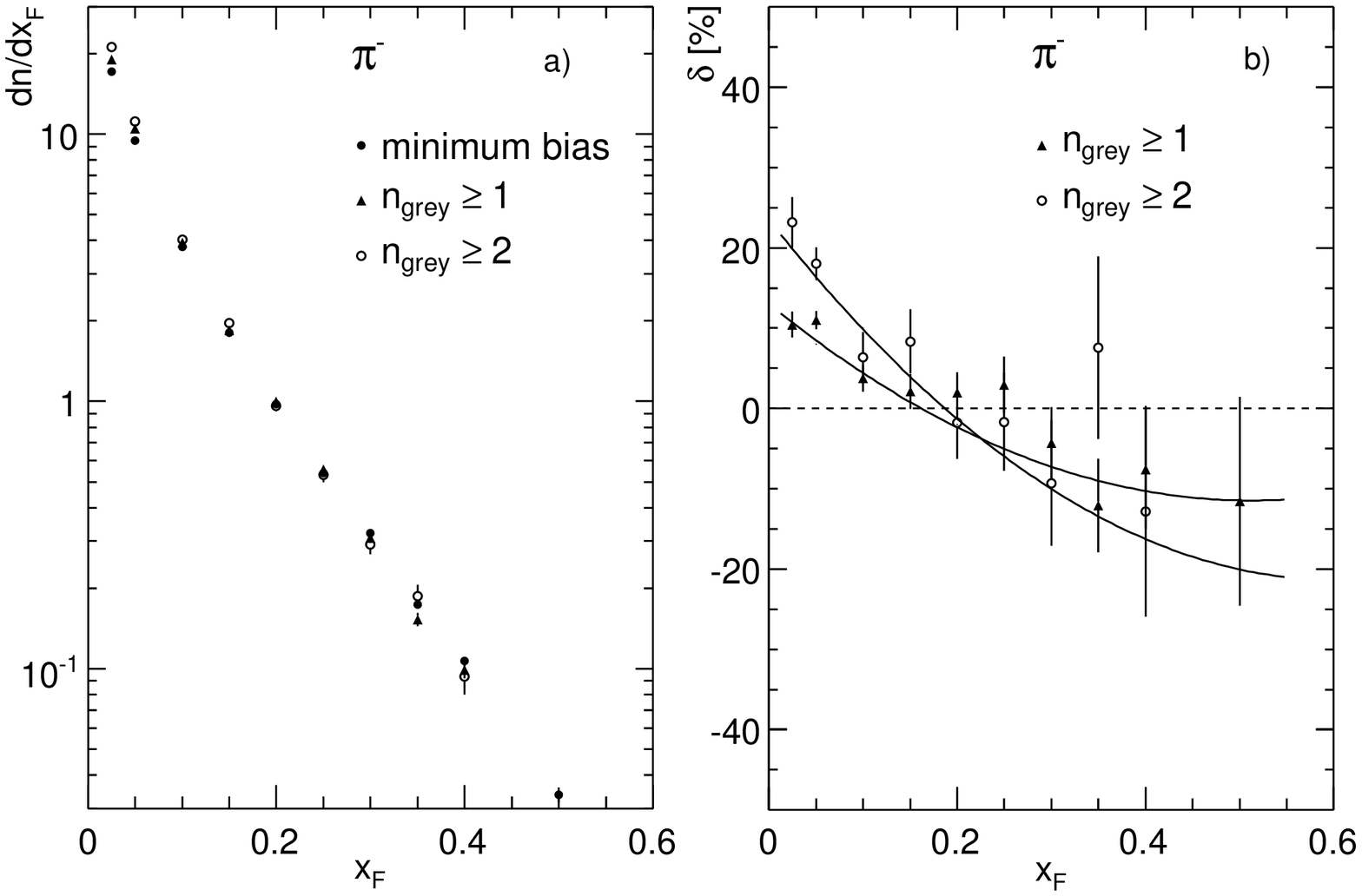,width=14cm}
  \caption{a) $dn/dx_F$ for $n_{\textrm{grey}}$ selection and b) the 
           difference $\delta$ between samples with $n_{\textrm{grey}}$ 
	   selection and minimum bias for $\pi^-$. 
	   The lines shown are to guide the eye}
  \label{fig:ngrey_neg}
\end{figure}

The extracted $p_T$ integrated pion yields are given in Table~\ref{tab:ngray}
for the total minimum bias sample and for the two centrality selections
$n_{\textrm{grey}}\geq$~1 and $n_{\textrm{grey}}\geq$~2. 
The distributions have been 
corrected by using the treatment developed in Sect.~\ref{sec:corr} 
integrated over transverse momentum.
The yields obtained for the minimum bias event sample are in good
statistical agreement with the values obtained from the integration of the 
interpolated double differential cross sections given in Table~\ref{tab:integr}.

In comparison to the $x_F$ dependence in minimum bias condition, a systematic
steepening is observed as shown in Figs.~\ref{fig:ngrey_pos}a and 
\ref{fig:ngrey_neg}a for $\pi^+$ and $\pi^-$, respectively. The difference 
$\delta(dn/dx_F) =( (dn/dx_F)_{n_{\textrm{grey}}} - 
(dn/dx_F)_{\textrm{min. bias}} )/(dn/dx_F)_{n_{\textrm{grey}}}$
is shown separately in Figs.~\ref{fig:ngrey_pos}b and \ref{fig:ngrey_neg}b.

%
%
\section{Availability of the presented data}
\vspace{3mm}                                                                                As in \cite{bib:pp_paper} the tabulated values of NA49 data are available in 
numerical form on the Web Site \cite{bib:site}. 
In addition two sets of $\pi^+$ and $\pi^-$ momentum vectors (5$\times$10$^7$ each) 
can be found on this site. They are generated via Monte Carlo following the
data interpolation presented in Sect.~\ref{sec:res} between the limits 
-0.1~$<x_F<$~0.5 and 0~$<p_T<$~2~GeV/c using a slight extrapolation 
of the data in the backward hemisphere at low $p_T$ and to $p_T$ values beyond 
1.8~GeV/c. Normalized invariant distributions may be deduced from these
vectors using the inelastic cross section given in Sect.~\ref{sec:trig}
and the following total integrated pion multiplicities in the
$x_F$ range from -0.1 to +0.5:
                                                           
\begin{align*}                     
  \langle n_{\pi^+}\rangle &= 3.279 \\
  \langle n_{\pi^-}\rangle &= 2.909 
\end{align*}

%
%
\section{Conclusions}
\vspace{3mm}
A new set of inclusive cross sections on pion production in
minimum bias p+C collisions at the CERN SPS is presented. The
data cover the central production region within the total range of 
-0.1~$<x_F<$~0.5 and 0~$<p_T<$~1.8~GeV/c for the first time.
The statistical uncertainties are typically at the few percent
level over the 270 measured bins per charge, with systematic
errors of less than 5\%. A detailed discussion of the results,
including in particular an in-depth comparison to the recently
published data from NA49 on p+p interactions, will be presented in
an accompanying paper.

\section*{Acknowledgements}
\vspace{3mm}
This work was supported by
the Bundesministerium f\"ur Bildung und Forschung, Germany,
the Polish State Committee for Scientific Research 
(1 P03B 006 30, SPB/CERN/P-03/Dz 446/2002-2004, 2 P03B 04123),
the Hungarian Scientific Research Foundation (T032648, T032293, T043514),
the Hungarian National Science Foundation, OTKA, (F034707),
the Polish-German Foundation,
the Bulgarian National Science Fund (Ph-09/05),
the EU FP6 HRM Marie Curie Intra-European Fellowship Program,
and 
the Particle Physics and Astronomy Research Council (PPARC) of the United Kingdom.

\end{document}